\documentclass{aa}

\usepackage[varg]{txfonts}

\usepackage{hyperref}
\usepackage{natbib}
\bibpunct{(}{)}{;}{a}{}{,} 

\def\msun{{\rm ~M}_{\odot}}

\def\mpy{{\rm ~M}_{\odot} {\rm ~yr}^{-1}}

\usepackage{color}

\begin{document}

   \title{Common origin for black holes in both high mass X-ray binaries and gravitational-wave sources}

   \author{Krzysztof Belczynski\inst{1}\textdagger
   \and 
   Christine Done\inst{2} 
   \and
   Scott Hagen\inst{2} 
   \and
   Jean-Pierre Lasota\inst{1,3} 
   \and 
   Koushik Sen\inst{4} 
          }

   \institute{Nicolaus Copernicus Astronomical Center, Polish Academy of Sciences,
          ul. Bartycka 18, 00-716 Warsaw, Poland             
         \and
             Department of Physics, University of Durham, South Road, Durham DH1 3LE, UK
         \and 
            Institut d'Astrophysique de Paris, CNRS et Sorbonne Universit\'e,
           UMR 7095, 98bis Bd Arago, 75014 Paris, France
         \and Institute of Astronomy, Faculty of Physics, Astronomy and Informatics, 
         Nicolaus Copernicus University, Grudzi{\c a}dzka 5, PL-87-100 Toru\'n, Poland
             }

             \offprints{J.-P. Lasota, \email{lasota@iap.fr}\\ \textdagger Deceased}

   \date{Received \today; accepted ...}

 
  \abstract
   {Black-hole (BH) high-mass X-ray binary (HMXB) systems are likely to be the progenitors of BH-BH mergers detected in gravitational waves by LIGO/Virgo/KAGRA (LVK). Yet merging BHs reach higher masses ($\sim 100\msun$) than BHs in HMXBs ($\sim 20\msun$) and typically exhibit lower spins ($a_{\rm BH}\lesssim 0.25$ with a larger values tail) than what is often claimed for BHs in HMXBs ($a_{\rm BH}\gtrsim 0.9$). This could suggest that these two classes of systems belong to different populations, but here we show that this may not necessarily be the case. The difference in masses is easily explained as the known HMXB-BHs are in galaxies with relatively high metallicity, so their progenitor stars are subject to strong mass loss from winds, leading to relatively low-mass BH at core collapse. Conversely, LVK is also able to detect BHs from low-metallicity galaxies that are known to naturally produce more massive stellar-origin BHs. {However, the difference in spin is more difficult to explain. Models with} efficient angular momentum transport in stellar interiors produce slowly spinning progenitors for both LVK and HMXB BHs. {Known} HMXBs have orbital periods that are too long for efficient tidal spin-up and are also unlikely to have undergone significant accretion spin-up. Instead, we show that  the derived value of the BH spin depends strongly on how the HMXB accretion disc emission is modelled. We argue that since Cyg X-1 is never observed to be in a soft spectral state, the appropriate spectral models must take into account the Comptonisation of the disc photosphere. We show that such models {are consistent with low spin values, namely: $a_{\rm BH}\sim 0.1$.} This was recently  confirmed  by other teams for both Cyg X-1 and LMC X-1 {and here we show this is also the case for M33 X-7. We conclude that all known HMXB BHs can exhibit a low spin}, in accordance with the results of stellar evolution models. Hence, the observations presented in this work are consistent with the LVK BHs and HMXB BHs belonging to the same population.
}

   \keywords{stars: black holes, compact objects, massive stars}

   \maketitle
   
   \nolinenumbers
   
%

\section{Introduction}
\label{sec.intro}

LIGO/Virgo/KAGRA (LVK) interferometers have detected gravitational waves from $\sim$ $70$ 
double black hole (BH-BH) mergers \citep{LigoO3b}. 
These merging black holes (BHs) have masses in the range $\sim 3-100\msun$, with many primary ({the} more 
massive) BHs having masses of $\sim 10\msun$ and $\sim 35\msun$\citep{Callister0424}. The most massive LVK event 
(GW190521) showed two merging BHs with estimated mass of $\sim 95\msun$ and $\sim 69\msun$.
The majority of these mergers have low positive effective spin parameters, with the distribution peaking at
\begin{equation} 
\chi_{\rm eff}= \frac{m_1 a_{\rm BH,1} \cos \theta_1+m_2 a_{\rm BH,2} \cos \theta_2}{m_1 + m_2} \approx 0.05, 
\end{equation}
where $m_{i}$ denotes BH masses, $a_{\rm BH,i}=cJ_{i}/Gm_{i}^2$ are the BH spin magnitudes ($J_{i}$: BH 
angular momentum, with $c \ $as the speed of light  and $\ G$ as the gravitational constant), while $\theta_{i}$ 
refers to the angles between the system's angular momentum and BH spins. Out of these $70$ mergers, $6$ of them show 
high positive effective spins $\chi_{\rm eff}>0.3$ (mean), while none show a high negative spin 
$\chi_{\rm eff}<-0.3$. However, there are three events with moderate negative spin estimates: 
$-0.3<\chi_{\rm eff}<-0.1$. Low values of the effective spin parameter may result from small 
individual BH spin magnitudes~\citep{Belczynski2020b}.  
This is supported by LVK data, which shows that BH individual spin magnitudes in BH-BH 
mergers peak at $a_{\rm BH} \sim 0.1-0.2$; however, the distribution has a long tail that extends to 
large values~(\citealt{LigoO3b}; {see \citealt{Callister0424} for slightly different results)}.

Electromagnetic observations have revealed a population of BHs hosted in binary star systems\footnote{e.g. see \url{https://universeathome.pl/universe/blackholes.php} and references therein}. Out of several 
known binary configurations hosting BHs, only high-mass X-ray binaries (HMXBs) can potentially lead 
to BH-BH mergers, as BH companion stars are massive enough to form a BH. {There are only three} HMXBs known
to have such massive ($\gtrsim 20\msun$) companions and {they have} moderate-mass BHs: { LMC X-1 with a $\sim 30\msun$ star and a $10.9\msun$ BH \citep{Orosz2009}; Cyg X-1 
with a $\sim 40\msun$ star and a $21.2\msun$ BH~\citep{MillerJones2021}; and M33 X-7 with a $\sim 38\msun$ star and a $11.4\msun$ BH (\citealt{Ramachandran2022}; {revised down from the previous estimates of a $70\msun$ donor and $15.6\msun$ BH in \citealt{Orosz2007}). We note that the binary LMC X-3 is not a HMBXB since its stellar companion is less massive ($3.63 \msun$) than the BH ($6.98\msun$) \citet{Orosz2014} and will not form a BH anyway.}}
The spins of BHs in HMXBs were obtained with two methods, namely: disc continuum fitting, {and reflection 
spectroscopy. Reflection spectroscopy is complex for soft state spectra (e.g. \citealt{Tomsick2018}), so here we focus on 
spin values from a disc continuum fitting. These are typically very high\footnote{{The spins attributed through various methods
to LMXB BH vary from very low to very high values, sometimes for the same system \citep[e.g.][]{Reynolds2021}. However, on the one hand these
binary systems are not BHBH binary progenitors, on the other, their evolutionary history is still a puzzle \citep{Wiktorowicz0914}.}}:
\begin{itemize}
\item Cyg X-1  $a_{\rm BH}> 0.95$ (e.g. \citealt{MillerJones2021,Gou2014} and 
    \item LMC X-1 $a_{\rm BH}=0.92$ \citep{Tripathi0720,Gou2009}
    \item M33 X-7 $a_{BH} =0.84$ from the old black hole mass  \cite{Liu2008,Liu2010}. However the newer, lower value
  gives $a_{\rm BH} = 0.7$ (this paper).
\end{itemize}
}

Since masses and spins of LVK BHs and HMXB BHs appear to be different, it 
might seem that these two groups of systems belong to populations with different formation scenarios. Here, we show 
that all these BHs may instead form a single population. The difference in BH mass arises naturally 
from the diverse formation environments: all known HMXB-BH systems reside in Local Group galaxies with relatively 
high metallicity, so they have strong stellar winds before the core collapse. This results in lower black 
hole masses compared to those forming in low metallicity galaxies, which are within the sensitivity 
range of LVK. We demonstrate the difference in expected black hole masses explicitly using a 
population synthesis model. This model also reproduces the spin values in the LVK BHs, for models where 
high spins are produced only by tidal spin-ups in extremely close binaries. The known HMXB-BH 
systems are in binary systems that are not close enough for tidal spin-up. Hence, the population synthesis models predict they should not have high spins, 
which conflicts with the reported high values. {However, the disc continuum method relies on the emission thermalising. If, instead, the disc is covered by a warm, optically thick skin, whereby the emission is Comptonised, the spin values are all consistent with $a_{\rm BH} \sim 0.1$ (\citealt{Belczynski1121,Zdziarski0224a,Zdziarski0224b}; this paper). With this interpretation,}  
the LVK BH-BH mergers and the HMXBs can be considered as members of 
the same binary star population. 

\section{BH masses}
\label{sec.mass}

To demonstrate the unicity of the BH population, we use 
the population synthesis code {\tt StarTrack}~\citep{Belczynski2002,Belczynski2008a}
and employ the delayed core-collapse supernova (SN) engine for neutron star/BH mass calculation
~\citep{Fryer2012} that does not produce a lower mass gap between NSs and BHs. 
{We used a model 
in which the
pair-instability (upper) mass gap is calculated with
non-standard fusion reaction rates and heavy mixing
for massive stars; this allows for BHs to form up to
$90\msun$ \citep{Farmer2020,Costa2021}. This type of model had already been developed by \citet{Belczynski2020c} in the context of the $M_{\rm BH}=90\msun$ detected at the low-metallicity GW190521 LVK merger, at 
distance of $\sim 4$~Gpc~\citep{gw190521a}. }

\begin{figure}
\hspace*{-0.4cm}
\includegraphics[width=0.5\textwidth]{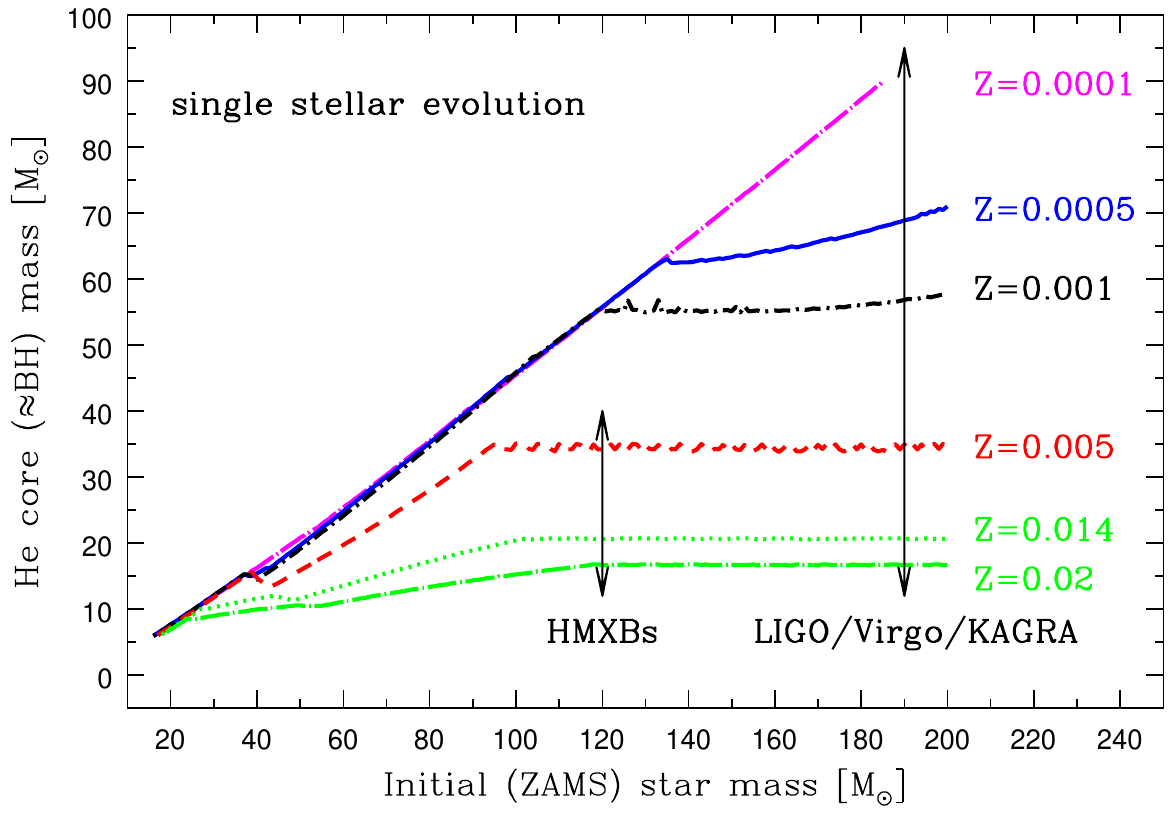}
\caption{
Helium core mass is a function of the initial star mass for various metallicities. 
The helium core is a good approximation of the black hole mass for stars in close 
binaries that form BH-BH mergers (binary interactions remove H-rich envelope). 
Black arrows show the current stellar metallicity range of LVK 
(low-redshift Universe: $z<0.7$) and for HMXBs that are limited to the local group of 
galaxies. LVK is bound to probe more massive BHs than electromagnetic HMXB  
observations.
}
\label{fig.bhmass}
\end{figure}

In Fig.~\ref{fig.bhmass}, we show the dependence of the helium core mass on metallicity for single 
stars obtained with the stellar wind treatment presented in~\cite{Belczynski2010b}. The helium core 
mass is a good approximation of the BH mass since binary interactions (Roche Lobe overflow or common 
envelope evolution) tend to remove the H-rich envelope from stars in close binaries (progenitors 
of BH-BH mergers if these form through isolated binary evolution). The lower the metallicity, 
the lower the opacity in stellar atmospheres, and the lower the wind mass loss, which produces progressively higher mass BHs with decreasing metallicity. 

Cyg X-1 is located in the Milky Way disc, which hosts stars that have metallicity values that are close to solar ($Z=0.014$,~\citealt{Asplund2009}). For such values, we obtain a maximum BH mass of 
$M_{\rm BH}\sim20\msun$, which is consistent with the highest mass stellar-origin BH known in the
Milky Way for the same Cyg X-1 ($M_{\rm BH}=21\msun$,~\citealt{MillerJones2021}). LMC X-1 is located in the 
Large Magellanic Cloud ($Z \approx 0.005$) and M33 X-7 in M33 galaxy ($Z=0.01$). In this metallicity 
range, BHs are not expected to be more massive than $M_{\rm BH}\sim35\msun$. 

LVK BH-BH mergers have already been detected (O3) to significant redshifts of $z \lesssim 0.7$. 
The detected BH-BH mergers may have formed even at much larger redshifts due to non-zero 
delay times between star formation and BH-BH mergers~\citep{Dominik2012,Fishbach2021b}. Therefore, 
galaxies and stars that can produce BH-BH mergers cover a wide range of metallicities. It is even 
claimed that Population III (metal-free) stars ($Z=0$) may contribute to LVK detections
~\citep{Kinugawa2014}. Assuming a metallicity range $Z=0.0001-0.02$ for LVK we obtain a
broad range of BH mass: $\sim 3-90\msun$. This range is consistent with the LVK BH mass 
estimates in BH-BH mergers. {The recent discovery of a $M_{\rm BH} \approx 33\msun$ BH in the binary Gaia BH3
\citep{Gaia33msun} strongly supports a scenario in which metal-poor massive stars 
are progenitors of the high-mass BHs observed by LVK \citep[see e.g.][]{Olejak2020a}.}
In conclusion, the same model of stellar evolution produces correct mass ranges for the three 
discussed HMXB BHs (narrow) and LVK BHs (wide). 

\section{BH spins: Stellar evolution}

Population synthesis models are also used to predict the BH spin distribution. In most cases, they produce 
low-spinning BHs ($a_{\rm BH}\sim 0.1$) from non-interacting stars and slow-to-rapidly-spinning 
BHs ($a_{\rm BH}=0.1-1.0$) for Wolf-Rayet (WR) stars that were subject to tidal spin-up in close 
binaries~\citep{Belczynski2020b,Bavera2020}. Such models can explain the low and high effective spins 
of LVK BH-BH mergers~\citep{Olejak2021b}. 

The low BH spins obtained by population synthesis models are the consequence of the assumed efficient angular
momentum transport from stellar cores to envelopes, an effect that has been now aptly confirmed by observations \citep{Aerts0819,Langer0912}
{consistent with the action} of the magnetic \citet{Tayler0173} instability \citep{Spruit0102,Fuller2019a,Eggenberger0822,Petitdemange0124}.
The heuristic \citet{Spruit0102} version of the mechanism used in population synthesis codes was recently confirmed by numerical calculations 
of \citet{Petitdemange0124}.

Therefore, unless their progenitors in HMXBs belong to a special category of stars with uncoupled cores and envelopes,
the BHs in these systems would have been formed with low spins  ($a_{\rm BH}\sim 0.1$). The {known} HMXB orbital periods are too long
for tidal interactions to spin up the stars before they form BHs {(the size of the present BH companion would
not fit into a $< 1.3$ days orbit of the tidal locking regime; see e.g. \citealt{Belczynski2020a}); thus,} 
 if their spins are really large, they {must have}
have gained angular momentum by accretion. To spin-up a BH from $a_{\rm BH}\sim 0.1$ to $\sim 0.9$ it is necessary to
 roughly to double its mass \citep{Bardeen0470}. For Cyg X-1, this would require accreting $\sim 10\msun$ in
less than $4$Myr \citep{MillerJones2021}, namely, with an accretion rate of $>2.5\times 10^{-6}\mpy$ {; this rate would be higher by factor of 100 if the
accretion lifetime of Cyg X-1 was only a few tens of thousands of years \citep{MillerJones2021}.} This gives an accretion rate of 
$> 6\dot M_{\rm Edd}$, where {$\dot M_{\rm Edd}=L_{Edd}/\eta c^2$ is the Eddington accretion rate with an assumed efficiency of $\sim 0.1$.}  Contrary to the still expressed opinion (see e.g. \citealt{MillerJones2021}),
the Eddington rate is not a maximum theoretical limit \citep{Begelman0478} and black holes can accrete from discs at rates of up
to (at least) a few tens of $\dot M_{\rm Edd}$ (\citealt{Kitaki0421,Yoshioka1222}, but see \citealt{Hu0822}); thus,
an accretion rate $ \gtrsim 12.5\dot M_{\rm Edd}$ in Cyg~X-1 would not be, from this point of view, extravagant.
The problem is to find an evolutionary path that would take a high-mass binary through such a super-Eddington phase
and produce the observed Cyg X-1, whose O supergiant donor star has yet to fill its Roche lobe. 
This is unlikely to happen with respect to accretion from wind, even when it is  L1-focused \citep{Wiktorowicz0721}. 
Therefore, according to the best stellar physics available BH spins in the known HMXBs should be low, with $a\lesssim 0.4$ -- consistent with LVK BH spins \citep{Fishbach2021a} and estimates of
BH spin in HMXBs by
\citet{Belczynski1121,Zdziarski0224a,Zdziarski0224b}; however, this is clearly incompatible with the generally accepted high spin values in these systems \citep{miller15}. 
These high spin values have been obtained by assuming the presence of a standard disc in the system (see Sect.\,\ref{sec.Discussion} for the standard-disc definition), which can be justified only
if the system is in a soft spectral state. In the following, we argue that since Cyg X-1 has never been observed in such a state, the {high black hole spin from disc continuum fitting is not 
a firm requirement.}

\section{BH spins from disc spectral fitting: HMXBs}
\label{seq.HMBXfit}

The main tracer of BH spin in electromagnetic systems is the radius of the innermost stable 
circular orbit ($R_{\rm ISCO}$). This sets the inner radius of the accretion disc in the standard 
Shakura-Sunyaev (\citealt{ss73} and its relativistic extension: Novikov-Thorne; \citealt{nt73}) models. 
Measuring the inner edge of the disc gives an estimate of BH spin, with $R_{\rm ISCO}$ decreasing 
from $6$ to $1.0 R_{\rm g}$ (with $R_{\rm g}$ being the gravitational radius) as $a_{\rm BH}$ increases 
from $0$ to $1.0$. For the \citet{Thorne0774} limit $a_{\rm BH}=0.998$, $R_{\rm ISCO}=1.24R_g$, but 
one should keep in mind that this limit is valid only for sub--Eddington flows (\citealt{Abramowicz0188,Sadowski0811}). The most direct way to measure the $R_{\rm ISCO}$ 
uses the luminosity emitted by the accretion disc itself.

Standard accretion disc models balance gravitational heating from viscous torques with blackbody 
cooling, resulting in an optically thick, geometrically thin structure with effective temperature at each 
radius, $T^4_{\rm eff}(R)\propto (L/R^2)(R_{\rm ISCO}/R)$. This gives a total spectrum, which is a sum 
of all radii of blackbody spectra of different temperatures. The emissivity peaks around 
$R\sim R_{\rm ISCO}$, so the peak temperature is $\propto (L/R_{\rm ISCO}^2)^{1/4}$~K, namely, it is a 
factor $2.2\,\times$ higher for the same disc luminosity at high spin than at low spin. 

In all X-ray binaries, we can observe cycles during which the brightness and spectral hardness form
a characteristic `tortoise-head' pattern. 
In general, the spectrum contains both thermal disc emission, together with a power law tail extending to much higher energies, 
clearly showing that there is an additional non-disc component to 
the accretion flow. Most transients show a phase in this cycle where the spectrum is
sufficiently soft to be termed `disc-dominated', where the accretion 
disc assumption is likely to be reliable. The problem is that not all X-ray sources complete
the full cycle. As we show below, this is the case for Cyg X-1.

\begin{figure}
\hspace*{-0.4cm}
\includegraphics[width=0.5\textwidth]{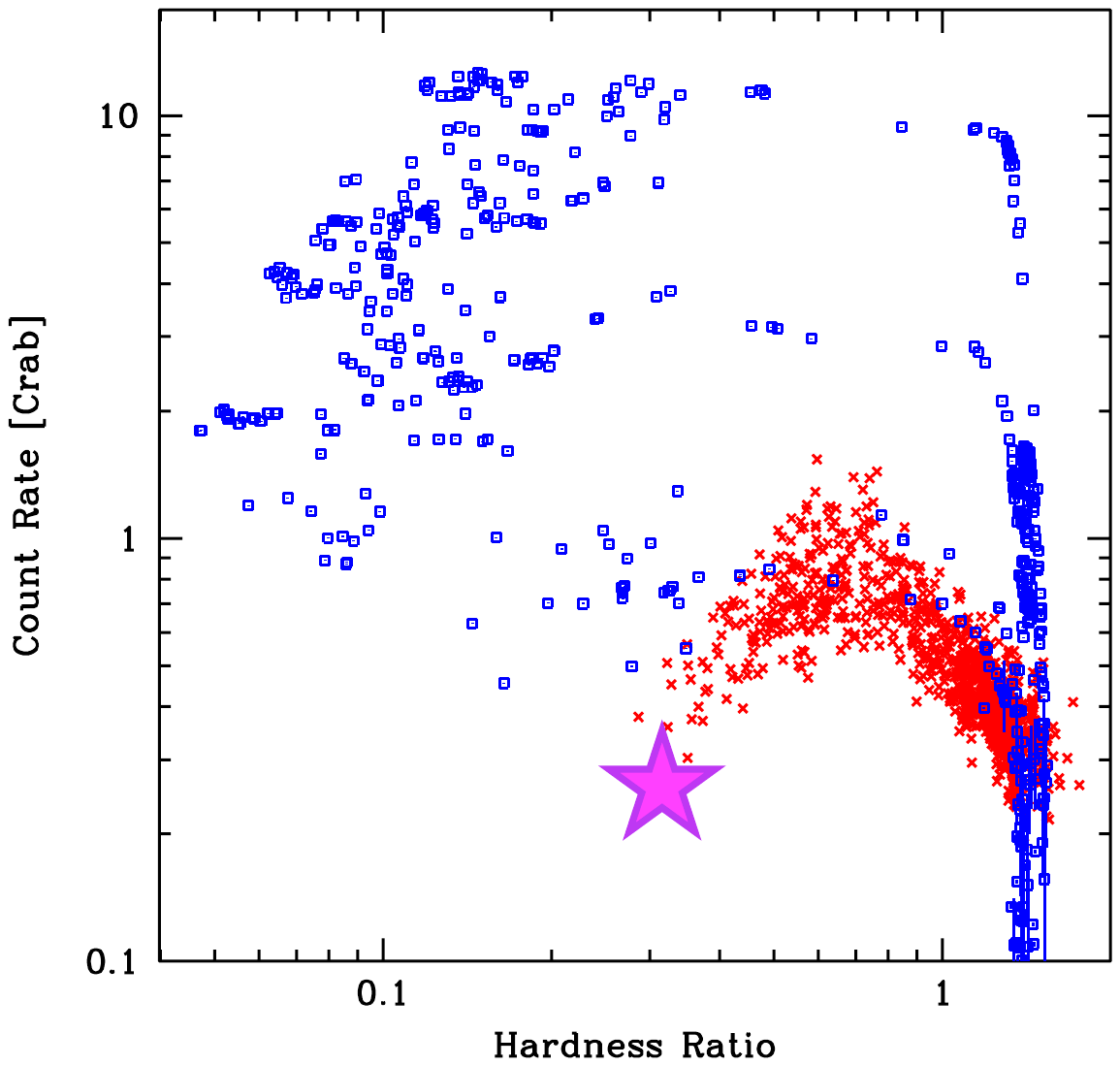}
\vspace{-40pt}
\caption{
The spectral hardness versus intensity diagram for all the RXTE data to 2005 from GX339-4 (blue) and Cyg X-1 (red), 
shifted in intensity so that the transition from soft to hard overlapped. Hardness ratio is defined as 
the ratio of counts in the energy bands $6.3 - 10.5$ and $3.8 - 6.3$\,keV. (Data from \citealt{bell10}.) {The Cyg X-1 Suzaku data used here is shown as the magenta star}. 
}
\label{fig.gx339_cyg}
\end{figure}

Indeed, there is clear evidence that in Cyg X-1, the accretion flow has not formed a 
standard blackbody disc. The blue points in Fig.~\ref{fig.gx339_cyg} show the hardness-intensity diagram of 
the LMXB GX339-4. 
This source picks out a well-defined track, transitioning from low and hard (bottom-right) to disc-dominated 
(upper left, hardness ratio below $\sim 0.2$), with fast transitions (horizontal) between the two states. The red points compare 
data from Cyg X-1 (both data taken from \citealt{bell10}). The mass and distance of GX339-4 are not 
well known, so we have scaled the relative count rate (brightness) between the two systems to match 
the lower transition luminosity. We can see how GX339-4 transitions from low/hard state 
(hardness ratio $>1$; the ratio is defined as the ratio of counts in the energy bands 
$6.3 - 10.5$ and $3.8 - 6.3$~keV) to high/soft state (hardness ratio of $<0.1$). In contrast, Cyg X-1 never 
reaches as far to the left as seen in the GX339-4 disc-dominated states. 

Hence, Cyg X-1 always has a strong tail of emission to higher energies (e.g. \citealt{gier99,Gou2009,wal16}), keeping it stuck in the more complex intermediate 
states where there is no reason to believe that the accretion flow forms a standard 
accretion disc. Instead, the soft  component in intermediate state spectra is often 
better fit by optically thick, warm (few keV) Comptonisation (see e.g. transition 
spectra from LMXB-BH MAXI J1820+070 in Fig.~2 of \citealt{Kawamura23}). 

Cyg X-1 has a higher mass BH ($21.2\msun$; \citealt{MillerJones2021}) than GX339-4 ($\sim 2-10\msun$; 
\citealt{Heida2017}), but this should make its disc spectra softer rather than harder since the ISCO is 
farther out for more massive BHs, making the disc temperature lower. 
A hotter disc temperature could instead result if Cyg X-1 has a substantially larger spin than GX339-4, 
but the reflection and reverberation spin estimates for GX339-4 are both extremely high. 
Therefore, there is no evidence that a standard disc forms in the whole available set of observations 
of Cyg X-1, making this system's standard disc-fitting method of BH spin determination inapplicable.

Instead, the strong soft component need not be emitted by a simple 
blackbody disc, where the energy is dissipated mainly on the midplane but can have heating further 
up into the photosphere. This forms a warm layer on the top of the disc, making it appear hotter and 
giving a higher spin when fit with standard disc models. Such a warm layer on the disc is 
sometimes required in the black hole binaries (e.g. in very high and intermediate type spectra: 
\citealt{kub01,Remillard2006}) and is generic in active galactic nuclei 
(\citealt{Porquet2019,Gierlinski2004b,petrucci18}). On the other hand,
it is well known that the simplistic assumption of dissipation that is concentrated at the disc
midplane (usually in the form of dissipation proportional to pressure) is in contradiction
with both observations (see e.g. \citealt{Hubeny0621}) and numerical simulations \citep{Blaes0914}.

Thus, there is no physical reason to assume that a standard accretion disc emits the observed X-ray spectrum of Cyg X-1. In fact, there are good reasons to think that such an assumption
is inappropriate for this system. There is no `canonical' model describing the inner disc
structure in X-ray binaries -- no model correctly describes the `tortoise-head' cycle. Hence, 
we  can test various options that would reproduce the observed spectrum and check whether
the deduced spin value is model-dependent; if it is, we consider which model corresponds best to
the value obtained from stellar evolution. A similar approach was adopted by \citet{Zdziarski0224a,Zdziarski0224b}.

\section{Cyg X-1 spin from the soft component}
\label{sec.cygx1}

\begin{figure*}
    \centering
    \hspace*{-1.6cm}
    \includegraphics[width=1.2\textwidth]{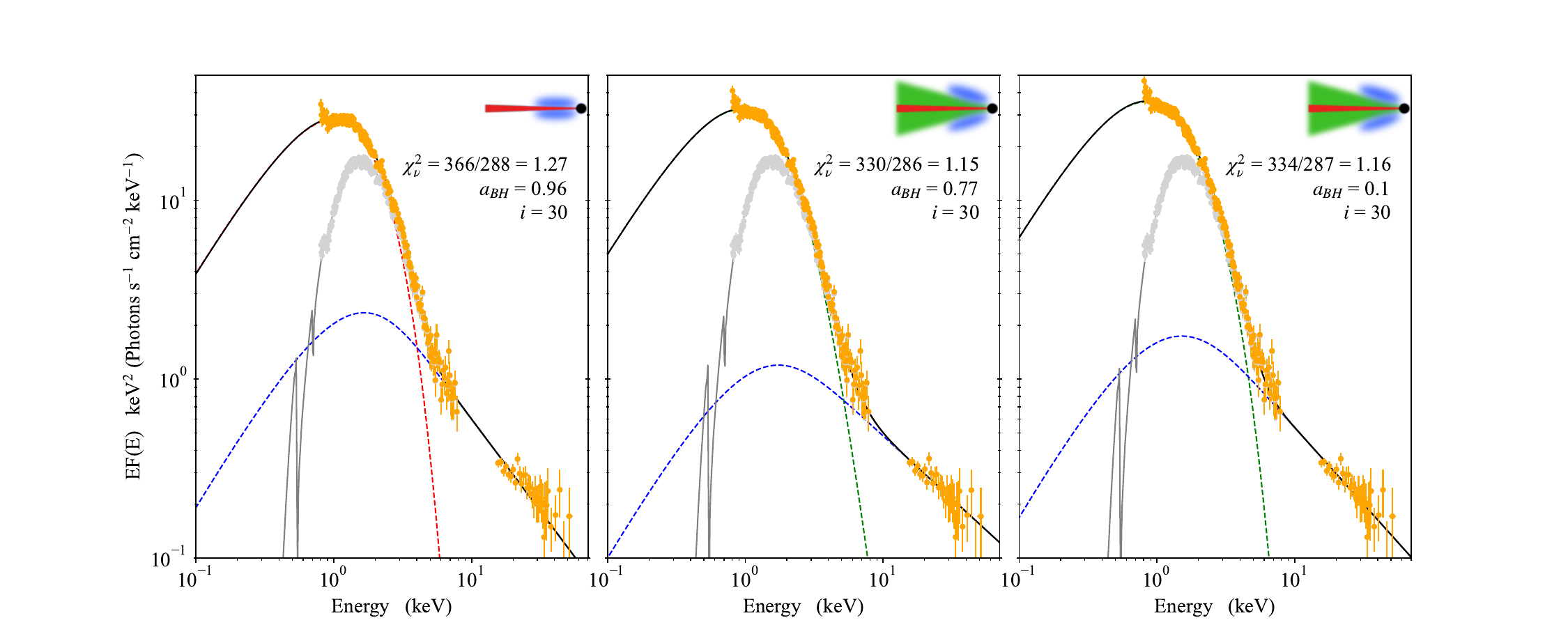}
    \caption{Cyg X-1 spectra for for: 
     {\it Left}: Standard accretion disc (red), modeled with {\sc kerrbb}, and a hot Comptonising 
     plasma (blue) giving the high energy emission - as sketched in the top right corner; model
     \textbf{(a)}.
    {\it Middle}: Standard accretion disc entirely covered by a warm Comptonising plasma (green), 
     and then an inner hot Comptonising plasma (blue); model \textbf{(b)}. 
    {\it Right}: Same as middle panel, but with the spin fixed at $a_{\rm BH}=0.1$.
    Note: the low BH spin is fully consistent with the X-ray spectral data.}
    \label{fig:cyg_3panel}
\end{figure*}

We confronted two models of Cyg X-1 emission.
We used data from the softest state ever seen (observed by {\sl Suzaku} in 2013; {magenta star in Fig. \ref{fig.gx339_cyg}}), when the soft component is at its most dominant, and, thus, uncertainties from modelling the high energy coronal tail are smallest (observation B4 from 
\citealt{Kawano2017}, see this paper for details of the data). We fit these data using the 
{\sc xspec} spectral fitting package and all the models below include absorption along the line of 
sight using {\sc tbabs} with column density as a free parameter {(see Appendix \ref{app.1} for the details)}. We show results from a series of 
model fits in Fig.~\ref{fig:cyg_3panel}, where the original (absorbed) X-ray Suzaku data is grey, 
with the reconstructed (corrected for absorption) spectrum shown with orange points. We assume here 
the standard mass and distance, and we fix the disc inclination to the binary inclination of
$i=30^\circ$. 

We first fit the standard disc model to these data. The results are shown in the left panel of 
Fig.~\ref{fig:cyg_3panel}. This includes all special and general relativistic ray tracing effects, 
as well as allowing a correction to the blackbody temperatures to account for the incomplete 
thermalisation expected from full models of the vertical disc structure (colour temperature 
correction, fixed to $1.7$). We allowed for a fraction of these disc spectra to be Compton upscattered to 
form the power law tail, modelled using the standard formalism, which assumes the electron energy is 
typically higher than $100$~keV. We recovered the usual, very high black hole spin 
($a_{\rm BH}=0.96$) seen in all standard-disc fits to the soft component (\citealt{Zhao0821,Zdziarski0224a,Zdziarski0224b}). 

The middle panel of Fig.~\ref{fig:cyg_3panel} shows what happens when we additionally allow a warm 
skin on the top of the disc. We model this using another Compton 
upscattering model, which allows the electron temperature to remain as a free parameter, as well as the 
scattered fraction (set by the optical depth of the skin). This is formally a much better fit to the 
data, even accounting for the two additional free parameters: optical depth and temperature of the 
warm disc skin. {A simple F-test gives a probability of the improvement being random as $p = 3.7 \times 10^{-7}$. This is well below the generally accepted limit of $p < 0.05$ for a result to be significant}. More importantly, the best fit reduces BH spin to $a_{\rm BH}=0.77$. The BH spin here is 
extremely poorly constrained as the soft component shape in this model is set more by the electron temperature 
of the skin than by the temperature of the disc at the ISCO. We illustrate this in the right panel 
of Fig.~\ref{fig:cyg_3panel}, where we show a fit with almost the same $\chi_\nu^2$ but for $a_{\rm BH}=0.1$, as predicted by standard stellar evolution.

This demonstrates that formal solutions for BH spin from spectra with their reported extremely small 
error bars or very definitive values (e.g. $a_{\rm BH}>0.9985$ at $3\sigma$ level,~\citealt{Zhao2021}) 
are model-dependent and are not the only solution (see also \citealt{Zdziarski0224a,Zdziarski0224b} who arrive at the same conclusion). 

\subsection{Considering whether a disc can form in Cyg X-1 for a low BH spin}
\label{sec.dis2}

The modelling of the disc continuum spectra assumes that an
accretion disc can form around the BH for any value of the BH 
spin and/or any orientation of the accretion disc with respect to
the spin of the BH. In a binary system, a disc will form only
if the circularisation radius of the transferred matter is larger
than the radius of the accreting body. In other words, a disc can 
form if the angular momentum of the accreting matter is large enough 
to allow it to circle around the accretor. For wind accretion (Bondi-Hoyle; 
\citealt{Bondi1952}),  the condition is \citep{FKR2002}:
\begin{equation}
    R_{\rm circ}=\frac{M_{\rm BH}^3\left(M_{\rm BH}+ M_*\right)}{16M_{*}^4}\left(\frac{R_*}{a}\right)^4 a \left(\frac{v_w}{v_{\rm esc}}\right)^{-8},
\end{equation}
where $M_*$ is the mass of the donor star, $a$ the orbital separation, 
$v_w$ the wind velocity, and $v_{\rm esc}$ the escape velocity from
the donor. For the parameters of Cyg X-1, when $a_{\rm BH}=0$, $R_{\rm ISCO}\approx 1.1 \times 10^7$cm but $R_{\rm circ} \approx 1.9 \times 10^{10} \left({v_w}/{v_{\rm esc}}\right)^{-8}$cm, so that, for instance, for ${v_w}/{v_{\rm esc}} > 2.2$ no disc can form.
\citet{Sen2021} analysed the problem in great detail and showed that for the
observed binary and stellar parameters of Cyg X-1 \citep{MillerJones2021},
a maximally spinning BH may be required to form an accretion disc 
in the prograde orientation {when the BH accretes from a wind}.

However, we note that the O star companion in Cyg X-1 is nearly
filling its Roche lobe \citep[Roche-lobe filling factor is 
greater than 0.9,][]{MillerJones2021},
where the formation of a focused accretion stream is expected
\citep{Blondin1991,Hadrava2012,ElMellah2019}. In Cyg X-1, the
existence of a focused accretion stream has also been observationally
verified \citep{Miller2005,Poutanen2008,Hanke2009}. {A high Roche 
lobe filling factor has been found for 
LMC X-1 \citep{Orosz2009} and the O donor in M33 X-7 effectively fills its Roche lobe 
(although not in a standard way; see \citealt{Ramachandran2022}). }

In such a configuration, spherically symmetric Bondi-Hoyle accretion
is not accurate. The tidal and gravitational effects of the BH distort 
the shape of the O star and wind streamlines from the O star, 
respectively \citep{ElMellah2019,Hirai2021}. \citet{Hirai2021} showed 
that for Cyg X-1, the effects of rotation and gravity darkening can 
lead to the formation of a prograde accretion disc around a non-rotating 
BH when the Roche lobe filling factor of the O star is greater than 
$\sim$0.8. 

Hence, the criterion of the formation of an accretion disc in Cyg
X-1 does not rule out any assumptions on the spin of the BH. The 
high Roche lobe filling factor of the O star enables the formation 
of a focused wind with sufficient angular momentum to make a prograde 
accretion disc. {In such cases, the circularisation radius for the bulk of the accretion flow
is of the order of the solar radius \citep[see e.g.][]{FKR2002}, orders
of magnitude larger than the ISCO radius}. The above arguments, together with our modelling 
of the disc continuum spectra, imply that the observations of Cyg\,X-1 
are consistent with the notion that its BH could also be slowly 
spinning.

\section{M33 X-7 BH spin from the soft component}
\label{sec.m33}

{The source X-7 in the nearby galaxy M33 is an eclipsing HMBH on a quasi-circular, 3.45 day orbit.
In this system, according to phase-resolved simultaneous HST and XMM-Newton observations by \citet{Ramachandran2022}, 
an $\approx 11.4\msun$ BH accretes matter lost from its O9II, metal-poor, $\approx 38 \msun$ supergiant companion.
The previous mass estimate by \citet{Orosz2007} had a higher BH mass of $\approx 15.6\msun$ (and $\approx 70 \msun$
companion). 

Assuming the older BH mass, \citet{Liu2008} used the continuum fitting method to get a BH spin $a_{\rm BH} = 0.84$.
We have used the standard continuum method to determine the M33 X-7 spin value for the new BH mass and
obtained $a_{\rm BH}= 0.7$ (see Appendix \ref{app.2}), as expected. With this new
spin value, the M33 X-7 BH is no longer 
in the very-rapidly rotating class, to which Cyg X-1
with $a_{\rm BH}\sim 0.9985$ and LMC X-1 with $a_{\rm BH}> 0.9$ belong. Nevertheless, since we have concluded that
the spin determination in Cyg X-1 is strongly model-dependent, supported by the findings of \citet{Zdziarski2023a,Zdziarski0224b}, who arrived to same conclusion
regarding LMC X-1, we thought it useful to check if the same applies to M33 X-7. Similarly to these sources, when a warm-skin Comptonising model
is used, the black hole spin is completely unconstrained, with acceptable values ranging from -1 to $\sim 0.9$ at $90\,\%$ confidence
level (see Appendix \ref{app.2}).
Below, we discuss the possible reason for this common property of the only three HMBXs that are potential progenitors of
BHBH binary systems.}

\section{Discussion}
\label{sec.Discussion}

{We have shown that the X-ray spectra of Cyg X-1 and M33 X-7  can be fitted with both standard disc and warm Comptonised disc
models (see also \citealt{Zdziarski0224a,Zdziarski0224b} for the same conclusion about Cyg X-1 and LMC X-1).}
We have also noticed that even in its softest state, the spectrum of Cyg X-1 is not soft 
enough to justify a standard-disc model fit, so there are very good reasons to consider
alternatives.

In order to avoid ambiguities and misunderstandings, one has to define precisely what is understood
by fitting spectra with a `standard disc' model. This model describes an axially symmetric, stationary, geometrically thin Keplerian accretion flow. It is assumed that the angular momentum transport is provided by a `viscous' (noting that in astrophysical discs, the viscosity
must be 'anomalous') torque and that the local dissipation of the same torque is the source of the energy
radiated away from the disc surface. Then, from the mass and angular momentum conservation equations for the
vertically averaged disc parameters, assuming thermal equilibrium, we obtain
\begin{equation}
\label{eq:disceq}
F=\sigma T_{\mathrm{eff}}^4 = \frac{3}{8 \pi} \frac{G M \dot{M}}{R^3}f,
\end{equation}
where $T_{\mathrm{eff}}$ is the effective temperature and $f$ contains information about the boundary conditions.
This is the equation that provides the flux -- or effective temperature profile, $T_{\mathrm{eff}}(R)$ 
-- for a Keplerian stationary disc. It does not say that the disc emits radiation like a blackbody. It
contains no information about the vertical energy transport in the disc and is independent of the angular momentum
and dissipation mechanism; in fact, it only assumes the existence of such mechanism (independently of the value of 
the viscosity parameter, $\alpha$, and does not even
assume the existence of such a parameter). Equation (\ref{eq:disceq}) is the basis of the `standard' disc X-ray spectrum
model. There, it is assumed that the colour temperature is close to the effective temperature; namely, the disc, locally radiates approximately as
a blackbody would. Since the accretor is a BH, one adds to the model general-relativistic effects, including radiation back-falling onto the disc,
and appropriate inner boundary conditions (usually zero torque at ISCO). Since the observed spectra are not strictly a sum of blackbodies, the disc annulus emission is represented by a colour-corrected (or diluted) blackbody:
\begin{equation}
I_\nu=B(f_cT)/f_c^4 = \frac{2 h}{c^2 f_c^4} \frac{\nu^3}{\exp \left(\frac{h \nu}{f_c k T}\right)-1} ,
\end{equation}
where $I_\nu$ is the intensity, $B_\nu$ is the Planck spectrum, $h$ the Planck constant, $\nu$ the radiation frequency, and $f_c$ is a {colour temperature correction factor
representing the extent to which 
electron scattering 
through the disc vertical structure 
is expected to affect the emergent spectrum.} This can be calculated by 
solving the disc vertical radiative transfer equations using a code such as TLUSTY
(see \citealt{Davis0319,Davis0305} and references therein). However, 
contrary 
to stellar atmospheres, accretion discs are not in radiative equilibrium, namely, the divergence of the radiative flux is not
equal to zero, but to the (viscous) heating rate per unit volume. Hence, the calculations
requires the viscous-dissipation vertical structure as input. 
In a simplistic application of the  $\alpha$-model of \citet{ss73}, it is assumed that this
dissipation profile is identical to that of the total pressure. Instead, for the disc emission calculations, the dissipation is assumed to follow the density profile in TLUSTY \citep{Done2008}. Both these prescriptions have 
maximal dissipation at the mid-plane, whereas MRI simulations often show
maximum dissipation closer to the disc surface \citep[see e.g.][]{Blaes0914}.
Some versions of TLUSTY \citep{Hubeny1098}
use a step-wise power-law dependence of viscosity that allows for a chosen fraction of the dissipation to occur in the surface layers.
Such a dissipation profile is needed to reproduce disc spectra of cataclysmic variables \cite{Hubeny0621} and when applied
to BH X-ray emitting disc, such dissipation stratification produces disc coronas \citep{Davis0319}, as suggested
by \citet{Svensson1294}.

In this regard, the real, physical dissipation profiles are still unknown since numerical simulations cannot yet produce realistic vertical
structures of geometrically thin accretion discs. That is why disc models with warm surface layers are as physically legitimate
as the presumed `standard' model, especially when there is strong evidence as in Cyg X-1 that the structure is not a standard disc. 

It is also useful to point out that the $\alpha$-disc model cannot explain the brightness-hardness ratio tortoise-head hysteresis observed 
in transient X-ray sources. The physical reason for this variability is still unknown. Observations (e.g. \citealt{Done2007,bas16,bas17}) 
and models \citep[e.g.][]{Dubus0701} suggests that accretion discs in BH X-ray binaries are truncated, but the physical mechanism of this excision
is still a mystery. 

This does not mean that the $\alpha$-disc model on which the `standard' spectral model is based is to be rejected. This model
has found many successful applications.
{The soft state spectra in transients show
peak disk temperature and total luminosity changing together with $L\propto T^{4}$, indicating a constant inner radius for the disc and constant $f_c$ \citep{Done2007}. 
The radial dependence of the disc temperature shows the predicted $T\propto R^{-3/4}$ behaviour in eclipsing cataclysmic variables in their disk (outbursting) states \citep{Horne0585}, as well as showing the very different radial profile expected from non-equilibrium discs in quiescent dwarf-novae; (\citealt{Lasota2001}). }

None of the known HMBHs exhibit a convincing $L \propto T^4$ soft state relation that characterises a 
standard disc (see Sect. \ref{seq.HMBXfit}).
Cyg X-1 is never in a soft state, while in
LMC X-1 \citet{Gierlinski2004} discovered that the disc
luminosity follows instead
a $L \propto T^{-1}$ track.  \citet{Zdziarski0224a} found that such a anti-correlation between luminosity and
temperature can be naturally explained as an effect of the warm Comptonisation layer on top of a standard disc. The luminosity variations in M33 X-7 are not large enough to determine whether a 
$L(T)$ relation is present \footnote{\citet{Liu2008} present a variability in {\sl counts} (not in counts per second)
which reflects mostly the variability of the exposure time.}.

The three BHHMBXs, Cyg X-1, LMC X-1, and M33 X-7, whose spectra are well fitted by a model of a disc covered with warm 
comptonising layer have one feature in common: they accrete not only from a focused stream, but also capture some matter
from the strong wind of the BH companion. 
It is natural to speculate that this is the source of the warm layer, making it a ubiquitous feature of the HMBH spectra.

\section{Conclusion}
\label{sec.summary}

The difference between HMXB and LVK BH masses has already been noticed after the first 
gravitational wave detection and attributed to metallicity effect (\citealt{AstroPaper}), as 
anticipated by \citet{Belczynski2010a}. Taking into account that interferometric detectors of 
gravitational waves sample a volume that is a few hundred thousand times larger than the X-ray 
telescopes (for HMXB BH mass determination), the difference in BH masses between LVK 
and HMXBs is not surprising as it implies a huge difference in the metallicity range sampled 
(see Fig.~\ref{fig.bhmass}).

Although assessing the problem of BH mass differences between HMXB and gravitational--wave sources is 
rather straightforward, the task of addressing the apparent conflicting spin values is more complicated. At 
face value, the spin values attributed to the BH in the HMXBs that are massive enough to form 
LVK sources (Cyg~X-~1, LMC X-1, M33 X-7) seem to require two distinct populations: slowly 
spinning (LVK) and rapidly spinning (HMXBs) BHs.

Since the majority of LVK BHs are slowly spinning ($a_{\rm BH}\sim 0.1-0.2$), we argue, under the assumption 
that they are stellar-origin BHs (i.e. they are not primordial BHs: \citealt{Hawking1971}, but see \citealt{Mroz0324}), that this must 
indicate that massive stars are subject to efficient angular momentum transport
~\citep{Spruit2002,Fuller2019a,Eggenberger0822,Petitdemange0124}. This leads to good agreement of predicted BH spins in BH-BH 
mergers with LVK data~\citep{Belczynski2020b,Bavera2020}. These slowly spinning BHs can form 
in isolated binary evolution and various dense (open, globular, nuclear) cluster 
environments. A small fraction of LVK BHs may be moderately or even rapidly spinning 
(see the long tail of BH spins with $a_{\rm BH}\gtrsim 0.4$: left panel of Fig.15 in \citealt{LigoO3b}). 
These moderately and rapidly spinning BHs can also be explained in the framework of efficient angular 
momentum transport. Although the star rotation slows down during its evolution towards the
formation of a BH, the tidal spin-up  of a WR star (BH progenitor) in close BH+WR binary systems 
(BH-BH progenitors) can easily result in a moderately or rapidly spinning BH~\citep{Olejak2021b}. 
In dense clusters, dynamical interactions may lead to the formation of 
BH-BH mergers with components that are second- or even third-generation BHs (formed by earlier BH-BH 
mergers) that naturally exhibit significant spins~\citep{Rodriguez2019a}.  

Under the natural assumption that BHs in HMXBs are formed directly from massive stars and within the 
paradigm of efficient angular momentum transport, these BHs also must have low spins ($a_{\rm BH}\sim 0.1$; e.g. 
~\citealt{Belczynski2020b}). We have demonstrated that this conclusion is fully consistent with
BH spin estimate based on a fitting of the disc spectra in HMXBs with models where the disc is covered by a
warm Comptonised layer. In particular, we show that with this model, the X-ray 
spectra of the softest (but still pretty hard) state of Cyg X-1 is consistent with a slowly spinning BH,
as confirmed by \citet{Zdziarski0224b} for another set of data. {The same is true of X-7 in M33, whose BH spin is completely unconstrained
when its soft X-ray spectra are fitted with the warm-skin model.}
The same warm-skin solutions also give a low 
spin for LMC X-1 (\citealt{Zdziarski0224a}. We have shown that the warm-skin solution is as physically plausible as the so--called
standard solution, especially in the case of systems that are not in a generic soft X-ray spectral
state.
We speculate that this warm skin is 
ubiquitous in the HMXB-BH due to the presence of additional accretion from
lower angular momentum material in the stellar wind as well as Roche 
Lobe overflow. 

Thus, we conclude that the LVK and HMXB BHs are consistent with having the same origin, that is, 
both classes of BHs form directly from stars with efficient angular momentum transport. 

\begin{acknowledgements}
      Unfortunately, Chris Belczyński died while working on the re-submission of the present paper. We will miss his deep knowledge, inexhaustible enthusiasm, and kindness.
      We thank Jerome Orosz, Henric Krawczynski, Andrew King, Aleksandra Olejak and Andrzej Zdziarski for very useful comments on this study. {This research has made use of data obtained from the \textit{Chandra} Data Archive and the \textit{Chandra} Source Catalog. We are grateful to the anonymous referee for very helpful suggestions and comments.}
      KB was supported by the Polish National Science Center (NCN) grant Maestro (2018/30/A/ST9/00050). CD acknowledges support from STFC through grant ST/P000541/1. SH acknowledges support from STFC through the studentship grant  ST/V506643/1. KS is funded by the National Science Center (NCN), Poland, under grant number OPUS 2021/41/B/ST9/00757.
\end{acknowledgements}

\bibliographystyle{aa} 
\bibliography{spin} 

\begin{thebibliography}{105}
\expandafter\ifx\csname natexlab\endcsname\relax\def\natexlab#1{#1}\fi

\bibitem[{{Abbott} {et~al.}(2016){Abbott}, {Abbott}, {Abbott}, {Abernathy},
  {Acernese}, {Ackley}, {Adams}, {Adams}, {Addesso}, {Adhikari},
  {et~al.}}]{AstroPaper}
{Abbott}, B.~P., {Abbott}, R., {Abbott}, T.~D., {et~al.} 2016, \apjl, 818, L22

\bibitem[{{Abbott} {et~al.}(2020){Abbott}, {Abbott}, {Abraham}, {Acernese},
  {Ackley}, {Adams}, {Adhikari}, {Adya}, {Affeldt}, {Agathos}, {Agatsuma},
  {Aggarwal}, {Aguiar}, {Aich}, {Aiello}, {Ain}, {Ajith}, {Akcay}, {Allen},
  {Allocca}, {LIGO Scientific Collaboration}, \& {Virgo
  Collaboration}}]{gw190521a}
{Abbott}, R., {Abbott}, T.~D., {Abraham}, S., {et~al.} 2020, \prl, 125, 101102

\bibitem[{{Abramowicz} \& {Lasota}(1980)}]{Abramowicz0188}
{Abramowicz}, M.~A. \& {Lasota}, J.~P. 1980, \actaa, 30, 35

\bibitem[{{Aerts} {et~al.}(2019){Aerts}, {Mathis}, \& {Rogers}}]{Aerts0819}
{Aerts}, C., {Mathis}, S., \& {Rogers}, T.~M. 2019, \araa, 57, 35

\bibitem[{{Arnaud}(1996)}]{Arnaud96}
{Arnaud}, K.~A. 1996, in Astronomical Society of the Pacific Conference Series,
  Vol. 101, Astronomical Data Analysis Software and Systems V, ed. G.~H.
  {Jacoby} \& J.~{Barnes}, 17

\bibitem[{{Asplund} {et~al.}(2009){Asplund}, {Grevesse}, {Sauval}, \&
  {Scott}}]{Asplund2009}
{Asplund}, M., {Grevesse}, N., {Sauval}, A.~J., \& {Scott}, P. 2009, \araa, 47,
  481

\bibitem[{{Bardeen}(1970)}]{Bardeen0470}
{Bardeen}, J.~M. 1970, \nat, 226, 64

\bibitem[{{Basak} \& {Zdziarski}(2016)}]{bas16}
{Basak}, R. \& {Zdziarski}, A.~A. 2016, \mnras, 458, 2199

\bibitem[{{Basak} {et~al.}(2017){Basak}, {Zdziarski}, {Parker}, \&
  {Islam}}]{bas17}
{Basak}, R., {Zdziarski}, A.~A., {Parker}, M., \& {Islam}, N. 2017, \mnras,
  472, 4220

\bibitem[{{Bavera} {et~al.}(2020){Bavera}, {Fragos}, {Qin}, {Zapartas},
  {Neijssel}, {Mandel}, {Batta}, {Gaebel}, {Kimball}, \&
  {Stevenson}}]{Bavera2020}
{Bavera}, S.~S., {Fragos}, T., {Qin}, Y., {et~al.} 2020, \aap, 635, A97

\bibitem[{{Begelman}(1979)}]{Begelman0478}
{Begelman}, M.~C. 1979, \mnras, 187, 237

\bibitem[{{Belczynski}(2020)}]{Belczynski2020c}
{Belczynski}, K. 2020, \apjl, 905, L15

\bibitem[{{Belczynski} {et~al.}(2010{\natexlab{a}}){Belczynski}, {Bulik},
  {Fryer}, {Ruiter}, {Valsecchi}, {Vink}, \& {Hurley}}]{Belczynski2010b}
{Belczynski}, K., {Bulik}, T., {Fryer}, C.~L., {et~al.} 2010{\natexlab{a}},
  \apj, 714, 1217

\bibitem[{{Belczynski} {et~al.}(2010{\natexlab{b}}){Belczynski}, {Dominik},
  {Bulik}, {O'Shaughnessy}, {Fryer}, \& {Holz}}]{Belczynski2010a}
{Belczynski}, K., {Dominik}, M., {Bulik}, T., {et~al.} 2010{\natexlab{b}},
  \apjl, 715, L138

\bibitem[{{Belczynski} {et~al.}(2021){Belczynski}, {Done}, {Hagen}, {Lasota},
  \& {Sen}}]{Belczynski1121}
{Belczynski}, K., {Done}, C., {Hagen}, S., {Lasota}, J.~P., \& {Sen}, K. 2021,
  arXiv e-prints, arXiv:2111.09401v2

\bibitem[{{Belczynski} {et~al.}(2020{\natexlab{a}}){Belczynski}, {Hirschi},
  {Kaiser}, {Liu}, {Casares}, {Lu}, {O'Shaughnessy}, {Heger}, {Justham}, \&
  {Soria}}]{Belczynski2020a}
{Belczynski}, K., {Hirschi}, R., {Kaiser}, E.~A., {et~al.} 2020{\natexlab{a}},
  \apj, 890, 113

\bibitem[{{Belczynski} {et~al.}(2002){Belczynski}, {Kalogera}, \&
  {Bulik}}]{Belczynski2002}
{Belczynski}, K., {Kalogera}, V., \& {Bulik}, T. 2002, \apj, 572, 407

\bibitem[{{Belczynski} {et~al.}(2008){Belczynski}, {Kalogera}, {Rasio}, {Taam},
  {Zezas}, {Bulik}, {Maccarone}, \& {Ivanova}}]{Belczynski2008a}
{Belczynski}, K., {Kalogera}, V., {Rasio}, F.~A., {et~al.} 2008, \apjs, 174,
  223

\bibitem[{{Belczynski} {et~al.}(2020{\natexlab{b}}){Belczynski}, {Klencki},
  {Fields}, {Olejak}, {Berti}, {Meynet}, {Fryer}, {Holz}, {O'Shaughnessy},
  {Brown}, {Bulik}, {Leung}, {Nomoto}, {Madau}, {Hirschi}, {Kaiser}, {Jones},
  {Mondal}, {Chruslinska}, {Drozda}, {Gerosa}, {Doctor}, {Giersz}, {Ekstrom},
  {Georgy}, {Askar}, {Baibhav}, {Wysocki}, {Natan}, {Farr}, {Wiktorowicz},
  {Coleman Miller}, {Farr}, \& {Lasota}}]{Belczynski2020b}
{Belczynski}, K., {Klencki}, J., {Fields}, C.~E., {et~al.} 2020{\natexlab{b}},
  \aap, 636, A104

\bibitem[{{Belloni}(2010)}]{bell10}
{Belloni}, T.~M. 2010, {States and Transitions in Black Hole Binaries}, ed.
  T.~{Belloni}, Vol. 794, 53

\bibitem[{{Blaes}(2014)}]{Blaes0914}
{Blaes}, O. 2014, \ssr, 183, 21

\bibitem[{{Blondin} {et~al.}(1991){Blondin}, {Stevens}, \&
  {Kallman}}]{Blondin1991}
{Blondin}, J.~M., {Stevens}, I.~R., \& {Kallman}, T.~R. 1991, \apj, 371, 684

\bibitem[{{Bondi}(1952)}]{Bondi1952}
{Bondi}, H. 1952, \mnras, 112, 195

\bibitem[{{Callister} \& {Farr}(2024)}]{Callister0424}
{Callister}, T.~A. \& {Farr}, W.~M. 2024, Physical Review X, 14, 021005

\bibitem[{{Costa} {et~al.}(2021){Costa}, {Bressan}, {Mapelli}, {Marigo},
  {Iorio}, \& {Spera}}]{Costa2021}
{Costa}, G., {Bressan}, A., {Mapelli}, M., {et~al.} 2021, \mnras, 501, 4514

\bibitem[{{Davis} {et~al.}(2005){Davis}, {Blaes}, {Hubeny}, \&
  {Turner}}]{Davis0305}
{Davis}, S.~W., {Blaes}, O.~M., {Hubeny}, I., \& {Turner}, N.~J. 2005, \apj,
  621, 372

\bibitem[{{Davis} \& {El-Abd}(2019)}]{Davis0319}
{Davis}, S.~W. \& {El-Abd}, S. 2019, \apj, 874, 23

\bibitem[{{Dominik} {et~al.}(2012){Dominik}, {Belczynski}, {Fryer}, {Holz},
  {Berti}, {Bulik}, {Mandel}, \& {O'Shaughnessy}}]{Dominik2012}
{Dominik}, M., {Belczynski}, K., {Fryer}, C., {et~al.} 2012, \apj, 759, 52

\bibitem[{{Done} \& {Davis}(2008)}]{Done2008}
{Done}, C. \& {Davis}, S.~W. 2008, \apj, 683, 389

\bibitem[{{Done} {et~al.}(2007){Done}, {Gierli{\'n}ski}, \&
  {Kubota}}]{Done2007}
{Done}, C., {Gierli{\'n}ski}, M., \& {Kubota}, A. 2007, \aapr, 15, 1

\bibitem[{{Dubus} {et~al.}(2001){Dubus}, {Hameury}, \& {Lasota}}]{Dubus0701}
{Dubus}, G., {Hameury}, J.~M., \& {Lasota}, J.~P. 2001, \aap, 373, 251

\bibitem[{{Eggenberger} {et~al.}(2022){Eggenberger}, {Moyano}, \& {den
  Hartogh}}]{Eggenberger0822}
{Eggenberger}, P., {Moyano}, F.~D., \& {den Hartogh}, J.~W. 2022, \aap, 664,
  L16

\bibitem[{{El Mellah} {et~al.}(2019){El Mellah}, {Sander}, {Sundqvist}, \&
  {Keppens}}]{ElMellah2019}
{El Mellah}, I., {Sander}, A.~A.~C., {Sundqvist}, J.~O., \& {Keppens}, R. 2019,
  \aap, 622, A189

\bibitem[{{Farmer} {et~al.}(2020){Farmer}, {Renzo}, {de Mink}, {Fishbach}, \&
  {Justham}}]{Farmer2020}
{Farmer}, R., {Renzo}, M., {de Mink}, S.~E., {Fishbach}, M., \& {Justham}, S.
  2020, \apjl, 902, L36

\bibitem[{{Fishbach} \& {Kalogera}(2021)}]{Fishbach2021b}
{Fishbach}, M. \& {Kalogera}, V. 2021, \apjl, 914, L30

\bibitem[{{Fishbach} \& {Kalogera}(2022)}]{Fishbach2021a}
{Fishbach}, M. \& {Kalogera}, V. 2022, \apjl, 929, L26

\bibitem[{{Frank} {et~al.}(2002){Frank}, {King}, \& {Raine}}]{FKR2002}
{Frank}, J., {King}, A., \& {Raine}, D.~J. 2002, {Accretion Power in
  Astrophysics: Third Edition} (CUP)

\bibitem[{{Fryer} {et~al.}(2012){Fryer}, {Belczynski}, {Wiktorowicz},
  {Dominik}, {Kalogera}, \& {Holz}}]{Fryer2012}
{Fryer}, C.~L., {Belczynski}, K., {Wiktorowicz}, G., {et~al.} 2012, \apj, 749,
  91

\bibitem[{{Fuller} {et~al.}(2019){Fuller}, {Piro}, \& {Jermyn}}]{Fuller2019a}
{Fuller}, J., {Piro}, A.~L., \& {Jermyn}, A.~S. 2019, \mnras, 485, 3661

\bibitem[{{Gaia Collaboration} {et~al.}(2024){Gaia Collaboration}, {Panuzzo},
  {Mazeh}, {Arenou}, {Holl}, {Caffau}, {Jorissen}, {Babusiaux}, {Gavras},
  {Sahlmann}, {Bastian}, {Wyrzykowski}, {Eyer}, {Leclerc}, {Bauchet},
  {Bombrun}, {Mowlavi}, {Seabroke}, {Teyssier}, {Balbinot}, {Helmi}, {Brown},
  {Vallenari}, {Prusti}, {de Bruijne}, {Barbier}, {Biermann}, {Creevey},
  {Ducourant}, {Evans}, {Guerra}, {Hutton}, {Jordi}, {Klioner}, {Lammers},
  {Lindegren}, {Luri}, {Mignard}, {Nicolas}, {Randich}, {Sartoretti},
  {Smiljanic}, {Tanga}, {Walton}, {Aerts}, {Bailer-Jones}, {Cropper},
  {Drimmel}, {Jansen}, {Katz}, {Lattanzi}, {Soubiran}, {Th{\'e}venin}, {van
  Leeuwen}, {Andrae}, {Audard}, {Bakker}, {Blomme}, {Casta{\~n}eda}, {De
  Angeli}, {Fabricius}, {Fouesneau}, {Fr{\'e}mat}, {Galluccio}, {Guerrier},
  {Heiter}, {Masana}, {Messineo}, {Nienartowicz}, {Pailler}, {Riclet}, {Roux},
  {Sordo}, {Gracia-Abril}, {Portell}, {Altmann}, {Benson}, {Berthier},
  {Burgess}, {Busonero}, {Busso}, {Cacciari}, {C{\'a}novas}, {Carrasco},
  {Carry}, {Cellino}, {Cheek}, {Clementini}, {Damerdji}, {Davidson}, {de
  Teodoro}, {Delchambre}, {Dell'Oro}, {Fraile Garcia}, {Garabato},
  {Garc{\'\i}a-Lario}, {Haigron}, {Hambly}, {Harrison}, {Hatzidimitriou},
  {Hern{\'a}ndez}, {Hestroffer}, {Hodgkin}, {Jamal}, {Jevardat de Fombelle},
  {Jordan}, {Krone-Martins}, {Lanzafame}, {L{\"o}ffler}, {Lorca}, {Marchal},
  {Marrese}, {Moitinho}, {Muinonen}, {Nu{\~n}ez Campos}, {Oreshina-Slezak},
  {Osborne}, {Pancino}, {Pauwels}, {Recio-Blanco}, {Riello}, {Rimoldini},
  {Robin}, {Roegiers}, {Sarro}, {Schultheis}, {Smith}, {Sozzetti}, {Utrilla},
  {van Leeuwen}, {Weingrill}, {Abbas}, {{\'A}brah{\'a}m}, {Abreu Aramburu},
  {Ahmed}, {Altavilla}, {{\'A}lvarez}, {Anders}, {Anderson}, {Anglada Varela},
  {Antoja}, {Baig}, {Baines}, {Baker}, {Balaguer-N{\'u}{\~n}ez}, {Balog},
  {Barache}, {Barros}, {Barstow}, {Bartolom{\'e}}, {Bashi}, {Bassilana},
  {Baudeau}, {Becciani}, {Bedin}, {Bellas-Velidis}, {Bellazzini}, {Beordo},
  {Bernet}, {Bertolotto}, {Bertone}, {Bianchi}, {Binnenfeld},
  {Blanco-Cuaresma}, {Bland-Hawthorn}, {Blazere}, {Boch}, {Bossini},
  {Bouquillon}, {Bragaglia}, {Braine}, {Bratsolis}, {Breedt}, {Bressan},
  {Brouillet}, {Brugaletta}, {Bucciarelli}, {Butkevich}, {Buzzi}, {Camut},
  {Cancelliere}, {Cantat-Gaudin}, {Capilla Guilarte}, {Carballo}, {Carlucci},
  {Carnerero}, {Carretero}, {Carton}, {Casamiquela}, {Casey}, {Castellani},
  {Castro-Ginard}, {Ceraj}, {Cesare}, {Charlot}, {Chaudet}, {Chemin},
  {Chiavassa}, {Chornay}, {Chosson}, {Cooper}, {Cornez}, {Cowell}, {Crosta},
  {Crowley}, {Cruz Reyes}, {Dafonte}, {Dal Ponte}, {David}, {de Laverny}, {De
  Luise}, {De March}, {De Ridder}, {de Torres}, {del Peloso}, {Delbo},
  {Delgado}, {Delisle}, {Demouchy}, {Denis}, {Dharmawardena}, {Di Giacomo},
  {Diener}, {Distefano}, {Dolding}, {Dsilva}, {Enke}, {Fabre}, {Fabrizio},
  {Faigler}, {Fatovi{\'c}}, {Fedorets}, {Fern{\'a}ndez-Hern{\'a}ndez},
  {Fernique}, {Figueras}, {Fouron}, {Fragkoudi}, {Gai}, {Galinier},
  {Garcia-Serrano}, {Garc{\'\i}a-Torres}, {Garofalo}, {Gerlach}, {Geyer},
  {Giacobbe}, {Gilmore}, {Girona}, {Giuffrida}, {Gomboc}, {Gomez},
  {Gonz{\'a}lez-Santamar{\'\i}a}, {Gosset}, {Granvik}, {Gregori Barrera},
  {Guti{\'e}rrez-S{\'a}nchez}, {Haywood}, {Helmer}, {Hidalgo}, {Hilger},
  {Hobbs}, {Hottier}, {Huckle}, {Jim{\'e}nez-Arranz}, {Juaristi Campillo},
  {Kaczmarek}, {Kervella}, {Khanna}, {Kontizas}, {Kordopatis}, {Korn},
  {K{\'o}sp{\'a}l}, {Kostrzewa-Rutkowska}, {Kruszy{\'n}ska}, {Kun}, {Lambert},
  {Lanza}, {Lebreton}, {Lebzelter}, {Leccia}, {Lecoutre}, {Liao}, {Liberato},
  {Licata}, {Livanou}, {Lobel}, {L{\'o}pez-Miralles}, {Loup}, {Madar{\'a}sz},
  {Mahy}, {Mann}, {Manteiga}, {Marinoni}, {Marcellino}, {Marshall},
  {Mascarenhas}, {Marchant}, {Mart{\'\i}n Lozano}, {Masip}, {Marconi},
  {Mar{\'\i}n Pina}, {Polo}, {Mart{\'\i}n-Fleitas}, {Mastrobuono-Battisti},
  {McMillan}, {Marton Meichsner}, {Merc}, {Messina}, {Millar}, {Mints},
  {Mohamed}, {Molina}, {Molinaro}, {Mongui{\'o}}, {Montegriffo}, {Monti},
  {Mora}, {Morbidelli}, {Morris}, {Mudimadugula}, {Muraveva}, {Musella},
  {Nagy}, {Nardetto}, {Navarrete}, {Oh}, {Ordenovic}, {Orenstein}, {Pagani},
  {Pagano}, {Palaversa}, {Palicio}, {Pallas-Quintela}, {Pawlak},
  {Penttil{\"a}}, {Pesciullesi}, {Pinamonti}, {Plachy}, {Planquart}, {Plum},
  {Poggio}, {Pourbaix}, {Price-Whelan}, {Pulone}, {Rabin}, {Rainer}, {Raiteri},
  {Ramos}, {Ramos-Lerate}, {Ratajczak}, {Re Fiorentin}, {Regibo}, {Reyl{\'e}},
  {Ripepi}, {Riva}, {Rix}, {Rixon}, {Robert}, {Robichon}, {Robin},
  {Romero-G{\'o}mez}, {Rowell}, {Ruz Mieres}, {Rybicki}, {Sadowski},
  {Sagrist{\`a} Sell{\'e}s}, {Sanna}, {Santove{\~n}a}, {Sarasso}, {Sarmiento},
  {Sarrate Riera}, {Sciacca}, {S{\'e}gransan}, {Semczuk}, {Shahaf}, {Siebert},
  {Slezak18}, {Smart}, {Snaith}, {Solano}, {Solitro}, {Souami}, {Souchay},
  {Spitoni}, {Spoto}, {Squillante}, {Steele}, {Steidelm{\"u}ller}, {Surdej},
  {Szabados}, {Taris}, {Taylor}, {Teixeira}, {Tepper-Garcia}, {Thuillot},
  {Tolomei}, {Tonello}, {Torra}, {Torralba Elipe}, {Trabucchi}, {Trentin},
  {Tsantaki}, {Turon}, {Ulla}, {Unger}, {Valtchanov}, {Vanel}, {Vecchiato},
  {Vicente}, {Villar}, {Weiler}, {Zhao}, {Zorec}, {Zucker}, {{\v{Z}}upi{\'c}},
  \& {Zwitter}}]{Gaia33msun}
{Gaia Collaboration}, {Panuzzo}, P., {Mazeh}, T., {et~al.} 2024, arXiv
  e-prints, arXiv:2404.10486

\bibitem[{{Gierli{\'n}ski} \& {Done}(2004{\natexlab{a}})}]{Gierlinski2004}
{Gierli{\'n}ski}, M. \& {Done}, C. 2004{\natexlab{a}}, \mnras, 347, 885

\bibitem[{{Gierli{\'n}ski} \& {Done}(2004{\natexlab{b}})}]{Gierlinski2004b}
{Gierli{\'n}ski}, M. \& {Done}, C. 2004{\natexlab{b}}, \mnras, 349, L7

\bibitem[{{Gierli{\'n}ski} {et~al.}(1999){Gierli{\'n}ski}, {Zdziarski},
  {Poutanen}, {Coppi}, {Ebisawa}, \& {Johnson}}]{gier99}
{Gierli{\'n}ski}, M., {Zdziarski}, A.~A., {Poutanen}, J., {et~al.} 1999,
  \mnras, 309, 496

\bibitem[{{Gou} {et~al.}(2009){Gou}, {McClintock}, {Liu}, {Narayan}, {Steiner},
  {Remillard}, {Orosz}, {Davis}, {Ebisawa}, \& {Schlegel}}]{Gou2009}
{Gou}, L., {McClintock}, J.~E., {Liu}, J., {et~al.} 2009, \apj, 701, 1076

\bibitem[{{Gou} {et~al.}(2014){Gou}, {McClintock}, {Remillard}, {Steiner},
  {Reid}, {Orosz}, {Narayan}, {Hanke}, \& {Garc{\'{\i}}a}}]{Gou2014}
{Gou}, L., {McClintock}, J.~E., {Remillard}, R.~A., {et~al.} 2014, \apj, 790,
  29

\bibitem[{{Hadrava} \& {{\v{C}}echura}(2012)}]{Hadrava2012}
{Hadrava}, P. \& {{\v{C}}echura}, J. 2012, \aap, 542, A42

\bibitem[{{Hanke} {et~al.}(2009){Hanke}, {Wilms}, {Nowak}, {Pottschmidt},
  {Schulz}, \& {Lee}}]{Hanke2009}
{Hanke}, M., {Wilms}, J., {Nowak}, M.~A., {et~al.} 2009, \apj, 690, 330

\bibitem[{{Hawking}(1971)}]{Hawking1971}
{Hawking}, S. 1971, \mnras, 152, 75

\bibitem[{{Heida} {et~al.}(2017){Heida}, {Jonker}, {Torres}, \&
  {Chiavassa}}]{Heida2017}
{Heida}, M., {Jonker}, P.~G., {Torres}, M.~A.~P., \& {Chiavassa}, A. 2017,
  \apj, 846, 132

\bibitem[{{Hirai} \& {Mandel}(2021)}]{Hirai2021}
{Hirai}, R. \& {Mandel}, I. 2021, PASA, 38, e056

\bibitem[{{Horne} \& {Cook}(1985)}]{Horne0585}
{Horne}, K. \& {Cook}, M.~C. 1985, \mnras, 214, 307

\bibitem[{{Hu} {et~al.}(2022){Hu}, {Inayoshi}, {Haiman}, {Li}, {Quataert}, \&
  {Kuiper}}]{Hu0822}
{Hu}, H., {Inayoshi}, K., {Haiman}, Z., {et~al.} 2022, \apj, 935, 140

\bibitem[{{Hubeny} \& {Hubeny}(1998)}]{Hubeny1098}
{Hubeny}, I. \& {Hubeny}, V. 1998, \apj, 505, 558

\bibitem[{{Hubeny} \& {Long}(2021)}]{Hubeny0621}
{Hubeny}, I. \& {Long}, K.~S. 2021, \mnras, 503, 5534

\bibitem[{{Kawamura} {et~al.}(2023){Kawamura}, {Done}, {Axelsson}, \&
  {Takahashi}}]{Kawamura23}
{Kawamura}, T., {Done}, C., {Axelsson}, M., \& {Takahashi}, T. 2023, \mnras,
  519, 4434

\bibitem[{{Kawano} {et~al.}(2017){Kawano}, {Done}, {Yamada}, {Takahashi},
  {Axelsson}, \& {Fukazawa}}]{Kawano2017}
{Kawano}, T., {Done}, C., {Yamada}, S., {et~al.} 2017, \pasj, 69, 36

\bibitem[{{Kinugawa} {et~al.}(2014){Kinugawa}, {Inayoshi}, {Hotokezaka},
  {Nakauchi}, \& {Nakamura}}]{Kinugawa2014}
{Kinugawa}, T., {Inayoshi}, K., {Hotokezaka}, K., {Nakauchi}, D., \&
  {Nakamura}, T. 2014, \mnras, 442, 2963

\bibitem[{{Kitaki} {et~al.}(2021){Kitaki}, {Mineshige}, {Ohsuga}, \&
  {Kawashima}}]{Kitaki0421}
{Kitaki}, T., {Mineshige}, S., {Ohsuga}, K., \& {Kawashima}, T. 2021, \pasj,
  73, 450

\bibitem[{{Kubota} {et~al.}(2001){Kubota}, {Makishima}, \& {Ebisawa}}]{kub01}
{Kubota}, A., {Makishima}, K., \& {Ebisawa}, K. 2001, \apjl, 560, L147

\bibitem[{{Langer}(2012)}]{Langer0912}
{Langer}, N. 2012, \araa, 50, 107

\bibitem[{{Lasota}(2001)}]{Lasota2001}
{Lasota}, J.-P. 2001, New Astronomy Reviews, 45, 449

\bibitem[{{Li} {et~al.}(2005){Li}, {Zimmerman}, {Narayan}, \&
  {McClintock}}]{Li05}
{Li}, L.-X., {Zimmerman}, E.~R., {Narayan}, R., \& {McClintock}, J.~E. 2005,
  \apjs, 157, 335

\bibitem[{{Liu} {et~al.}(2008){Liu}, {McClintock}, {Narayan}, {Davis}, \&
  {Orosz}}]{Liu2008}
{Liu}, J., {McClintock}, J.~E., {Narayan}, R., {Davis}, S.~W., \& {Orosz},
  J.~A. 2008, \apjl, 679, L37

\bibitem[{{Liu} {et~al.}(2010){Liu}, {McClintock}, {Narayan}, {Davis}, \&
  {Orosz}}]{Liu2010}
{Liu}, J., {McClintock}, J.~E., {Narayan}, R., {Davis}, S.~W., \& {Orosz},
  J.~A. 2010, \apjl, 719, L109

\bibitem[{{Miller} {et~al.}(2005){Miller}, {Wojdowski}, {Schulz}, {Marshall},
  {Fabian}, {Remillard}, {Wijnands}, \& {Lewin}}]{Miller2005}
{Miller}, J.~M., {Wojdowski}, P., {Schulz}, N.~S., {et~al.} 2005, \apj, 620,
  398

\bibitem[{{Miller} \& {Miller}(2015)}]{miller15}
{Miller}, M.~C. \& {Miller}, J.~M. 2015, \physrep, 548, 1

\bibitem[{{Miller-Jones} {et~al.}(2021){Miller-Jones}, {Bahramian}, {Orosz},
  {Mandel}, {Gou}, {Maccarone}, {Neijssel}, {Zhao}, {Zi{\'o}{\l}kowski},
  {Reid}, {Uttley}, {Zheng}, {Byun}, {Dodson}, {Grinberg}, {Jung}, {Kim},
  {Marcote}, {Markoff}, {Rioja}, {Rushton}, {Russell}, {Sivakoff}, {Tetarenko},
  {Tudose}, \& {Wilms}}]{MillerJones2021}
{Miller-Jones}, J. C.~A., {Bahramian}, A., {Orosz}, J.~A., {et~al.} 2021,
  Science, 371, 1046

\bibitem[{{Mr{\'o}z} {et~al.}(2024){Mr{\'o}z}, {Udalski}, {Szyma{\'n}ski},
  {Soszy{\'n}ski}, {Wyrzykowski}, {Pietrukowicz}, {Koz{\l}owski}, {Poleski},
  {Skowron}, {Skowron}, {Ulaczyk}, {Gromadzki}, {Rybicki}, {Iwanek}, {Wrona},
  \& {Ratajczak}}]{Mroz0324}
{Mr{\'o}z}, P., {Udalski}, A., {Szyma{\'n}ski}, M.~K., {et~al.} 2024, \nat,
  632, 749

\bibitem[{{Novikov} \& {Thorne}(1973)}]{nt73}
{Novikov}, I.~D. \& {Thorne}, K.~S. 1973, in Black Holes (Les Astres Occlus),
  343--450

\bibitem[{{Olejak} \& {Belczynski}(2021)}]{Olejak2021b}
{Olejak}, A. \& {Belczynski}, K. 2021, \apjl, 921, L2

\bibitem[{{Olejak} {et~al.}(2020){Olejak}, {Belczynski}, {Bulik}, \&
  {Sobolewska}}]{Olejak2020a}
{Olejak}, A., {Belczynski}, K., {Bulik}, T., \& {Sobolewska}, M. 2020, \aap,
  638, A94

\bibitem[{{Orosz} {et~al.}(2007){Orosz}, {McClintock}, {Narayan}, {Bailyn},
  {Hartman}, {Macri}, {Liu}, {Pietsch}, {Remillard}, {Shporer}, \&
  {Mazeh}}]{Orosz2007}
{Orosz}, J.~A., {McClintock}, J.~E., {Narayan}, R., {et~al.} 2007, \nat, 449,
  872

\bibitem[{{Orosz} {et~al.}(2009){Orosz}, {Steeghs}, {McClintock}, {Torres},
  {Bochkov}, {Gou}, {Narayan}, {Blaschak}, {Levine}, {Remillard}, {Bailyn},
  {Dwyer}, \& {Buxton}}]{Orosz2009}
{Orosz}, J.~A., {Steeghs}, D., {McClintock}, J.~E., {et~al.} 2009, \apj, 697,
  573

\bibitem[{{Orosz} {et~al.}(2014){Orosz}, {Steiner}, {McClintock}, {Buxton},
  {Bailyn}, {Steeghs}, {Guberman}, \& {Torres}}]{Orosz2014}
{Orosz}, J.~A., {Steiner}, J.~F., {McClintock}, J.~E., {et~al.} 2014, \apj,
  794, 154

\bibitem[{{Petitdemange} {et~al.}(2024){Petitdemange}, {Marcotte}, {Gissinger},
  \& {Daniel}}]{Petitdemange0124}
{Petitdemange}, L., {Marcotte}, F., {Gissinger}, C., \& {Daniel}, F. 2024,
  \aap, 681, A75

\bibitem[{{Petrucci} {et~al.}(2018){Petrucci}, {Ursini}, {De Rosa}, {Bianchi},
  {Cappi}, {Matt}, {Dadina}, \& {Malzac}}]{petrucci18}
{Petrucci}, P.~O., {Ursini}, F., {De Rosa}, A., {et~al.} 2018, \aap, 611, A59

\bibitem[{{Porquet} {et~al.}(2019){Porquet}, {Done}, {Reeves}, {Grosso},
  {Marinucci}, {Matt}, {Lobban}, {Nardini}, {Braito}, {Marin}, {Kubota},
  {Ricci}, {Koss}, {Stern}, {Ballantyne}, \& {Farrah}}]{Porquet2019}
{Porquet}, D., {Done}, C., {Reeves}, J.~N., {et~al.} 2019, \aap, 623, A11

\bibitem[{{Poutanen} {et~al.}(2008){Poutanen}, {Zdziarski}, \&
  {Ibragimov}}]{Poutanen2008}
{Poutanen}, J., {Zdziarski}, A.~A., \& {Ibragimov}, A. 2008, \mnras, 389, 1427

\bibitem[{{Ramachandran} {et~al.}(2022){Ramachandran}, {Oskinova}, {Hamann},
  {Sander}, {Todt}, {Pauli}, {Shenar}, {Torrej{\'o}n}, {Postnov}, {Blondin},
  {Bozzo}, {Hainich}, \& {Massa}}]{Ramachandran2022}
{Ramachandran}, V., {Oskinova}, L.~M., {Hamann}, W.~R., {et~al.} 2022, \aap,
  667, A77

\bibitem[{{Remillard} \& {McClintock}(2006)}]{Remillard2006}
{Remillard}, R.~A. \& {McClintock}, J.~E. 2006, \araa, 44, 49

\bibitem[{{Reynolds}(2021)}]{Reynolds2021}
{Reynolds}, C.~S. 2021, \araa, 59, 117

\bibitem[{{Rodriguez} {et~al.}(2019){Rodriguez}, {Zevin}, {Amaro-Seoane},
  {Chatterjee}, {Kremer}, {Rasio}, \& {Ye}}]{Rodriguez2019a}
{Rodriguez}, C.~L., {Zevin}, M., {Amaro-Seoane}, P., {et~al.} 2019, \prd, 100,
  043027

\bibitem[{{Sen} {et~al.}(2021){Sen}, {Xu}, {Langer}, {El Mellah},
  {Sch{\"u}rmann}, \& {Quast}}]{Sen2021}
{Sen}, K., {Xu}, X.~T., {Langer}, N., {et~al.} 2021, \aap, 652, A138

\bibitem[{{Shakura} \& {Sunyaev}(1973)}]{ss73}
{Shakura}, N.~I. \& {Sunyaev}, R.~A. 1973, \aap, 500, 33

\bibitem[{{S{\k{a}}dowski} {et~al.}(2011){S{\k{a}}dowski}, {Bursa},
  {Abramowicz}, {Klu{\'z}niak}, {Lasota}, {Moderski}, \&
  {Safarzadeh}}]{Sadowski0811}
{S{\k{a}}dowski}, A., {Bursa}, M., {Abramowicz}, M., {et~al.} 2011, \aap, 532,
  A41

\bibitem[{{Spruit}(2002{\natexlab{a}})}]{Spruit0102}
{Spruit}, H.~C. 2002{\natexlab{a}}, \aap, 381, 923

\bibitem[{{Spruit}(2002{\natexlab{b}})}]{Spruit2002}
{Spruit}, H.~C. 2002{\natexlab{b}}, \aap, 381, 923

\bibitem[{{Steiner} {et~al.}(2009){Steiner}, {Narayan}, {McClintock}, \&
  {Ebisawa}}]{Steiner09}
{Steiner}, J.~F., {Narayan}, R., {McClintock}, J.~E., \& {Ebisawa}, K. 2009,
  \pasp, 121, 1279

\bibitem[{{Svensson} \& {Zdziarski}(1994)}]{Svensson1294}
{Svensson}, R. \& {Zdziarski}, A.~A. 1994, \apj, 436, 599

\bibitem[{{Tayler}(1973)}]{Tayler0173}
{Tayler}, R.~J. 1973, \mnras, 161, 365

\bibitem[{{The LIGO Scientific Collaboration} {et~al.}(2023){The LIGO
  Scientific Collaboration}, {the Virgo Collaboration}, {the KAGRA
  Collaboration}, {Abbott}, {Abbott}, {Acernese}, {Ackley}, {Adams},
  {Adhikari}, {Adhikari}, {Adya}, {Affeldt}, {Agarwal}, {Agathos}, {Agatsuma},
  {Aggarwal}, {Aguiar}, {Aiello}, {Ain}, {Ajith}, {Akutsu}, {Albanesi},
  {Allocca}, {Altin}, {Amato}, {Anand}, {Anand}, {Ananyeva}, {Anderson},
  {Anderson}, {Ando}, {Andrade}, {Andres}, {Andri{\'c}}, {Angelova}, {Ansoldi},
  {Antelis}, {Antier}, {Antonini}, {Appert}, {Arai}, {Arai}, {Arai}, {Araki},
  {Araya}, {Araya}, {Areeda}, {Ar{\`e}ne}, {Aritomi}, {Arnaud}, {Aronson},
  {Arun}, {Asada}, {Asali}, {Ashton}, {Aso}, {Assiduo}, {Aston}, {Astone},
  {Aubin}, {Austin}, {Babak}, {Badaracco}, {Bader}, {Badger}, {Bae}, {Bae},
  {Baer}, {Bagnasco}, {Bai}, {Baiotti}, {Baird}, {Bajpai}, {Ball}, {Ballardin},
  {Ballmer}, {Balsamo}, {Baltus}, {Banagiri}, {Bankar}, {Barayoga}, {Barbieri},
  {Barish}, {Barker}, {Barneo}, {Barone}, {Barr}, {Barsotti}, {Barsuglia},
  {Barta}, {Bartlett}, {Barton}, {Bartos}, {Bassiri}, {Basti}, {Bawaj},
  {Bayley}, {Baylor}, {Bazzan}, {B{\'e}csy}, {Bedakihale}, {Bejger},
  {Belahcene}, {Benedetto}, {Beniwal}, {Bennett}, {Bentley}, {BenYaala},
  {Bergamin}, {Berger}, {Bernuzzi}, {Berry}, {Bersanetti}, {Bertolini},
  {Betzwieser}, {Beveridge}, {Bhandare}, {Bhardwaj}, {Bhattacharjee},
  {Bhaumik}, {Bilenko}, {Billingsley}, {Bini}, {Birney}, {Birnholtz},
  {Biscans}, {Bischi}, {Biscoveanu}, {Bisht}, {Biswas}, {Bitossi}, {Bizouard},
  {Blackburn}, {Blair}, {Blair}, {Blair}, {Bobba}, {Bode}, {Boer}, {Bogaert},
  {Boldrini}, {Bonavena}, {Bondu}, {Bonilla}, {Bonnand}, {Booker}, {Boom},
  {Bork}, {Boschi}, {Bose}, {Bose}, {Bossilkov}, {Boudart}, {Bouffanais},
  {Bozzi}, {Bradaschia}, {Brady}, {Bramley}, {Branch}, {Branchesi}, {Brau},
  {Breschi}, {Briant}, {Briggs}, {Brillet}, {Brinkmann}, {Brockill}, {Brooks},
  {Brooks}, {Brown}, {Brunett}, {Bruno}, {Bruntz}, {Bryant}, {Bulik}, {Bulten},
  {Buonanno}, {Buscicchio}, {Buskulic}, {Buy}, {Byer}, {Cadonati}, {Cagnoli},
  {Cahillane}, {Calder{\'o}n Bustillo}, {Callaghan}, {Callister}, {Calloni},
  {Cameron}, {Camp}, {Canepa}, {Canevarolo}, {Cannavacciuolo}, {Cannon}, {Cao},
  {Cao}, {Capocasa}, {Capote}, {Carapella}, {Carbognani}, {Carlin}, {Carney},
  {Carpinelli}, {Carrillo}, {Carullo}, {Carver}, {Casanueva Diaz}, {Casentini},
  {Castaldi}, {Caudill}, {Cavagli{\`a}}, {Cavalier}, {Cavalieri}, {Ceasar},
  {Cella}, {Cerd{\'a}-Dur{\'a}n}, {Cesarini}, {Chaibi}, {Chakravarti},
  {Chalathadka Subrahmanya}, {Champion}, {Chan}, {Chan}, {Chan}, {Chan},
  {Chan}, {Chandra}, {Chanial}, {Chao}, {Charlton}, {Chase},
  {Chassande-Mottin}, {Chatterjee}, {Chatterjee}, {Chatterjee}, {Chaturvedi},
  {Chaty}, {Chatziioannou}, {Chen}, {Chen}, {Chen}, {Chen}, {Chen}, {Chen},
  {Chen}, {Chen}, {Cheng}, {Cheong}, {Cheung}, {Chia}, {Chiadini}, {Chiang},
  {Chiarini}, {Chierici}, {Chincarini}, {Chiofalo}, {Chiummo}, {Cho}, {Cho},
  {Choudhary}, {Choudhary}, {Christensen}, {Chu}, {Chu}, {Chu}, {Chua},
  {Chung}, {Ciani}, {Ciecielag}, {Cie{\'s}lar}, {Cifaldi}, {Ciobanu}, {Ciolfi},
  {Cipriano}, {Cirone}, {Clara}, {Clark}, {Clark}, {Clarke}, {Clearwater},
  {Clesse}, {Cleva}, {Coccia}, {Codazzo}, {Cohadon}, {Cohen}, {Cohen},
  {Colleoni}, {Collette}, {Colombo}, {Colpi}, {Compton}, {Constancio}, {Conti},
  {Cooper}, {Corban}, {Corbitt}, {Cordero-Carri{\'o}n}, {Corezzi}, {Corley},
  {Cornish}, {Corre}, {Corsi}, {Cortese}, {Costa}, {Cotesta}, {Coughlin},
  {Coulon}, {Countryman}, {Cousins}, {Couvares}, {Coward}, {Cowart}, {Coyne},
  {Coyne}, {Creighton}, {Creighton}, {Criswell}, {Croquette}, {Crowder},
  {Cudell}, {Cullen}, {Cumming}, {Cummings}, {Cunningham}, {Cuoco},
  {Cury{\l}o}, {Dabadie}, {Dal Canton}, {Dall'Osso}, {D{\'a}lya}, {Dana},
  {DaneshgaranBajastani}, {D'Angelo}, {Danilishin}, {D'Antonio}, {Danzmann},
  {Darsow-Fromm}, {Dasgupta}, {Datrier}, {Datta}, {Dattilo}, {Dave}, {Davier},
  {Davies}, {Davis}, {Davis}, {Daw}, {Dean}, {DeBra}, {Deenadayalan},
  {Degallaix}, {De Laurentis}, {Del{\'e}glise}, {Del Favero}, {De Lillo}, {De
  Lillo}, {Del Pozzo}, {DeMarchi}, {De Matteis}, {D'Emilio}, {Demos}, {Dent},
  {Depasse}, {De Pietri}, {De Rosa}, {De Rossi}, {DeSalvo}, {De Simone},
  {Dhurandhar}, {D{\'\i}az}, {Diaz-Ortiz}, {Didio}, {Dietrich}, {Di Fiore}, {Di
  Fronzo}, {Di Giorgio}, {Di Giovanni}, {Di Giovanni}, {Di Girolamo}, {Di
  Lieto}, {Ding}, {Di Pace}, {Di Palma}, {Di Renzo}, {Divakarla}, {Dmitriev},
  {Doctor}, {D'Onofrio}, {Donovan}, {Dooley}, {Doravari}, {Dorrington},
  {Drago}, {Driggers}, {Drori}, {Ducoin}, {Dupej}, {Durante}, {D'Urso},
  {Duverne}, {Dwyer}, {Eassa}, {Easter}, {Ebersold}, {Eckhardt}, {Eddolls},
  {Edelman}, {Edo}, {Edy}, {Effler}, {Eguchi}, {Eichholz}, {Eikenberry},
  {Eisenmann}, {Eisenstein}, {Ejlli}, {Engelby}, {Enomoto}, {Errico}, {Essick},
  {Estell{\'e}s}, {Estevez}, {Etienne}, {Etzel}, {Evans}, {Evans}, {Ewing},
  {Fafone}, {Fair}, {Fairhurst}, {Farah}, {Farinon}, {Farr}, {Farr}, {Farrow},
  {Fauchon-Jones}, {Favaro}, {Favata}, {Fays}, {Fazio}, {Feicht}, {Fejer},
  {Fenyvesi}, {Ferguson}, {Fernandez-Galiana}, {Ferrante}, {Ferreira},
  {Fidecaro}, {Figura}, {Fiori}, {Fishbach}, {Fisher}, {Fittipaldi}, {Fiumara},
  {Flaminio}, {Floden}, {Fong}, {Font}, {Fornal}, {Forsyth}, {Franke},
  {Frasca}, {Frasconi}, {Frederick}, {Freed}, {Frei}, {Freise}, {Frey},
  {Fritschel}, {Frolov}, {Fronz{\'e}}, {Fujii}, {Fujikawa}, {Fukunaga},
  {Fukushima}, {Fulda}, {Fyffe}, {Gabbard}, {Gadre}, {Gair}, {Gais},
  {Galaudage}, {Gamba}, {Ganapathy}, {Ganguly}, {Gao}, {Gaonkar}, {Garaventa},
  {Garc{\'\i}a-N{\'u}{\~n}ez}, {Garc{\'\i}a-Quir{\'o}s}, {Garufi}, {Gateley},
  {Gaudio}, {Gayathri}, {Ge}, {Gemme}, {Gennai}, {George}, {Gerberding},
  {Gergely}, {Gewecke}, {Ghonge}, {Ghosh}, {Ghosh}, {Ghosh}, {Ghosh},
  {Giacomazzo}, {Giacoppo}, {Giaime}, {Giardina}, {Gibson}, {Gier}, {Giesler},
  {Giri}, {Gissi}, {Glanzer}, {Gleckl}, {Godwin}, {Goetz}, {Goetz}, {Gohlke},
  {Golomb}, {Goncharov}, {Gonz{\'a}lez}, {Gopakumar}, {Gosselin}, {Gouaty},
  {Gould}, {Grace}, {Grado}, {Granata}, {Granata}, {Grant}, {Gras}, {Grassia},
  {Gray}, {Gray}, {Greco}, {Green}, {Green}, {Gretarsson}, {Gretarsson},
  {Griffith}, {Griffiths}, {Griggs}, {Grignani}, {Grimaldi}, {Grimm}, {Grote},
  {Grunewald}, {Gruning}, {Guerra}, {Guidi}, {Guimaraes}, {Guix{\'e}},
  {Gulati}, {Guo}, {Guo}, {Gupta}, {Gupta}, {Gupta}, {Gustafson}, {Gustafson},
  {Guzman}, {Ha}, {Haegel}, {Hagiwara}, {Haino}, {Halim}, {Hall}, {Hamilton},
  {Hammond}, {Han}, {Haney}, {Hanks}, {Hanna}, {Hannam}, {Hannuksela},
  {Hansen}, {Hansen}, {Hanson}, {Harder}, {Hardwick}, {Haris}, {Harms},
  {Harry}, {Harry}, {Hartwig}, {Hasegawa}, {Haskell}, {Hasskew}, {Haster},
  {Hattori}, {Haughian}, {Hayakawa}, {Hayama}, {Hayes}, {Healy}, {Heidmann},
  {Heidt}, {Heintze}, {Heinze}, {Heinzel}, {Heitmann}, {Hellman}, {Hello},
  {Helmling-Cornell}, {Hemming}, {Hendry}, {Heng}, {Hennes}, {Hennig},
  {Hennig}, {Hernandez}, {Hernandez Vivanco}, {Heurs}, {Hild}, {Hill},
  {Himemoto}, {Hines}, {Hiranuma}, {Hirata}, {Hirose}, {Hochheim}, {Hofman},
  {Hohmann}, {Holcomb}, {Holland}, {Hollows}, {Holmes}, {Holt}, {Holz}, {Hong},
  {Hopkins}, {Hough}, {Hourihane}, {Howell}, {Hoy}, {Hoyland}, {Hreibi},
  {Hsieh}, {Hsu}, {Huang}, {Huang}, {Huang}, {Huang}, {Huang}, {Huang},
  {H{\"u}bner}, {Huddart}, {Hughey}, {Hui}, {Hui}, {Husa}, {Huttner},
  {Huxford}, {Huynh-Dinh}, {Ide}, {Idzkowski}, {Iess}, {Ikenoue}, {Imam},
  {Inayoshi}, {Ingram}, {Inoue}, {Ioka}, {Isi}, {Isleif}, {Ito}, {Itoh},
  {Iyer}, {Izumi}, {JaberianHamedan}, {Jacqmin}, {Jadhav}, {Jadhav}, {James},
  {Jan}, {Jani}, {Janquart}, {Janssens}, {Janthalur}, {Jaranowski}, {Jariwala},
  {Jaume}, {Jenkins}, {Jenner}, {Jeon}, {Jeunon}, {Jia}, {Jin}, {Johns},
  {Jones}, {Jones}, {Jones}, {Jones}, {Jones}, {Jonker}, {Ju}, {Jung}, {Jung},
  {Junker}, {Juste}, {Kaihotsu}, {Kajita}, {Kakizaki}, {Kalaghatgi},
  {Kalogera}, {Kamai}, {Kamiizumi}, {Kanda}, {Kandhasamy}, {Kang}, {Kanner},
  {Kao}, {Kapadia}, {Kapasi}, {Karat}, {Karathanasis}, {Karki}, {Kashyap},
  {Kasprzack}, {Kastaun}, {Katsanevas}, {Katsavounidis}, {Katzman}, {Kaur},
  {Kawabe}, {Kawaguchi}, {Kawai}, {Kawasaki}, {K{\'e}f{\'e}lian}, {Keitel},
  {Key}, {Khadka}, {Khalili}, {Khan}, {Khazanov}, {Khetan}, {Khursheed},
  {Kijbunchoo}, {Kim}, {Kim}, {Kim}, {Kim}, {Kim}, {Kim}, {Kimball}, {Kimura},
  {Kinley-Hanlon}, {Kirchhoff}, {Kissel}, {Kita}, {Kitazawa}, {Kleybolte},
  {Klimenko}, {Knee}, {Knowles}, {Knyazev}, {Koch}, {Koekoek}, {Kojima},
  {Kokeyama}, {Koley}, {Kolitsidou}, {Kolstein}, {Komori}, {Kondrashov},
  {Kong}, {Kontos}, {Koper}, {Korobko}, {Kotake}, {Kovalam}, {Kozak},
  {Kozakai}, {Kozu}, {Kringel}, {Krishnendu}, {Kr{\'o}lak}, {Kuehn}, {Kuei},
  {Kuijer}, {Kumar}, {Kumar}, {Kumar}, {Kumar}, {Kume}, {Kuns}, {Kuo}, {Kuo},
  {Kuromiya}, {Kuroyanagi}, {Kusayanagi}, {Kuwahara}, {Kwak}, {Lagabbe},
  {Laghi}, {Lalande}, {Lam}, {Lamberts}, {Landry}, {Landry}, {Lane}, {Lang},
  {Lange}, {Lantz}, {La Rosa}, {Lartaux-Vollard}, {Lasky}, {Laxen},
  {Lazzarini}, {Lazzaro}, {Leaci}, {Leavey}, {Lecoeuche}, {Lee}, {Lee}, {Lee},
  {Lee}, {Lee}, {Lee}, {Lehmann}, {Lema{\^\i}tre}, {Leonardi}, {Leroy},
  {Letendre}, {Levesque}, {Levin}, {Leviton}, {Leyde}, {Li}, {Li}, {Li}, {Li},
  {Li}, {Li}, {Lin}, {Lin}, {Lin}, {Lin}, {Lin}, {Linde}, {Linker}, {Linley},
  {Littenberg}, {Liu}, {Liu}, {Liu}, {Liu}, {Llamas}, {Llorens-Monteagudo},
  {Lo}, {Lockwood}, {London}, {Longo}, {Lopez}, {Lopez Portilla}, {Lorenzini},
  {Loriette}, {Lormand}, {Losurdo}, {Lott}, {Lough}, {Lousto}, {Lovelace},
  {Lucaccioni}, {L{\"u}ck}, {Lumaca}, {Lundgren}, {Luo}, {Lynam}, {Macas},
  {MacInnis}, {Macleod}, {MacMillan}, {Macquet}, {Maga{\~n}a Hernandez},
  {Magazz{\`u}}, {Magee}, {Maggiore}, {Magnozzi}, {Mahesh}, {Majorana},
  {Makarem}, {Maksimovic}, {Maliakal}, {Malik}, {Man}, {Mandic}, {Mangano},
  {Mango}, {Mansell}, {Manske}, {Mantovani}, {Mapelli}, {Marchesoni},
  {Marchio}, {Marion}, {Mark}, {M{\'a}rka}, {M{\'a}rka}, {Markakis},
  {Markosyan}, {Markowitz}, {Maros}, {Marquina}, {Marsat}, {Martelli},
  {Martin}, {Martin}, {Martinez}, {Martinez}, {Martinez}, {Martinovic},
  {Martynov}, {Marx}, {Masalehdan}, {Mason}, {Massera}, {Masserot},
  {Massinger}, {Masso-Reid}, {Mastrogiovanni}, {Matas}, {Mateu-Lucena},
  {Matichard}, {Matiushechkina}, {Mavalvala}, {McCann}, {McCarthy},
  {McClelland}, {McClincy}, {McCormick}, {McCuller}, {McGhee}, {McGuire},
  {McIsaac}, {McIver}, {McRae}, {McWilliams}, {Meacher}, {Mehmet}, {Mehta},
  {Meijer}, {Melatos}, {Melchor}, {Mendell}, {Menendez-Vazquez}, {Menoni},
  {Mercer}, {Mereni}, {Merfeld}, {Merilh}, {Merritt}, {Merzougui}, {Meshkov},
  {Messenger}, {Messick}, {Meyers}, {Meylahn}, {Mhaske}, {Miani}, {Miao},
  {Michaloliakos}, {Michel}, {Michimura}, {Middleton}, {Milano}, {Miller},
  {Miller}, {Miller}, {Miller}, {Millhouse}, {Mills}, {Milotti}, {Minazzoli},
  {Minenkov}, {Mio}, {Mir}, {Miravet-Ten{\'e}s}, {Mishra}, {Mishra}, {Mistry},
  {Mitra}, {Mitrofanov}, {Mitselmakher}, {Mittleman}, {Miyakawa}, {Miyamoto},
  {Miyazaki}, {Miyo}, {Miyoki}, {Mo}, {Moguel}, {Mogushi}, {Mohapatra},
  {Mohite}, {Molina}, {Molina-Ruiz}, {Mondin}, {Montani}, {Moore}, {Moraru},
  {Morawski}, {More}, {Moreno}, {Moreno}, {Mori}, {Morisaki}, {Moriwaki},
  {Mours}, {Mow-Lowry}, {Mozzon}, {Muciaccia}, {Mukherjee}, {Mukherjee},
  {Mukherjee}, {Mukherjee}, {Mukherjee}, {Mukund}, {Mullavey}, {Munch},
  {Mu{\~n}iz}, {Murray}, {Musenich}, {Muusse}, {Nadji}, {Nagano}, {Nagano},
  {Nagar}, {Nakamura}, {Nakano}, {Nakano}, {Nakashima}, {Nakayama}, {Napolano},
  {Nardecchia}, {Narikawa}, {Naticchioni}, {Nayak}, {Nayak}, {Negishi}, {Neil},
  {Neilson}, {Nelemans}, {Nelson}, {Nery}, {Neubauer}, {Neunzert}, {Ng}, {Ng},
  {Nguyen}, {Nguyen}, {Nguyen}, {Nguyen Quynh}, {Ni}, {Nichols}, {Nishizawa},
  {Nissanke}, {Nitoglia}, {Nocera}, {Norman}, {North}, {Nozaki}, {Nuttall},
  {Oberling}, {O'Brien}, {Obuchi}, {O'Dell}, {Oelker}, {Ogaki}, {Oganesyan},
  {Oh}, {Oh}, {Oh}, {Ohashi}, {Ohishi}, {Ohkawa}, {Ohme}, {Ohta}, {Okada},
  {Okutani}, {Okutomi}, {Olivetto}, {Oohara}, {Ooi}, {Oram}, {O'Reilly},
  {Ormiston}, {Ormsby}, {Ortega}, {O'Shaughnessy}, {O'Shea}, {Oshino},
  {Ossokine}, {Osthelder}, {Otabe}, {Ottaway}, {Overmier}, {Pace}, {Pagano},
  {Page}, {Pagliaroli}, {Pai}, {Pai}, {Palamos}, {Palashov}, {Palomba}, {Pan},
  {Pan}, {Panda}, {Pang}, {Pang}, {Pankow}, {Pannarale}, {Pant}, {Panther},
  {Paoletti}, {Paoli}, {Paolone}, {Parisi}, {Park}, {Park}, {Parker},
  {Pascucci}, {Pasqualetti}, {Passaquieti}, {Passuello}, {Patel}, {Pathak},
  {Patricelli}, {Patron}, {Paul}, {Payne}, {Pedraza}, {Pegoraro}, {Pele},
  {Pe{\~n}a Arellano}, {Penn}, {Perego}, {Pereira}, {Pereira}, {Perez},
  {P{\'e}rigois}, {Perkins}, {Perreca}, {Perri{\`e}s}, {Petermann},
  {Petterson}, {Pfeiffer}, {Pham}, {Phukon}, {Piccinni}, {Pichot},
  {Piendibene}, {Piergiovanni}, {Pierini}, {Pierro}, {Pillant}, {Pillas},
  {Pilo}, {Pinard}, {Pinto}, {Pinto}, {Piotrzkowski}, {Pirello}, {Pitkin},
  {Placidi}, {Planas}, {Plastino}, {Pluchar}, {Poggiani}, {Polini}, {Pong},
  {Ponrathnam}, {Popolizio}, {Porter}, {Poulton}, {Powell}, {Pracchia},
  {Pradier}, {Prajapati}, {Prasai}, {Prasanna}, {Pratten}, {Principe}, {Prodi},
  {Prokhorov}, {Prosposito}, {Prudenzi}, {Puecher}, {Punturo}, {Puosi},
  {Puppo}, {P{\"u}rrer}, {Qi}, {Quetschke}, {Quitzow-James}, {Raab},
  {Raaijmakers}, {Radkins}, {Radulesco}, {Raffai}, {Rail}, {Raja}, {Rajan},
  {Ramirez}, {Ramirez}, {Ramos-Buades}, {Rana}, {Rapagnani}, {Rapol}, {Ray},
  {Raymond}, {Raza}, {Razzano}, {Read}, {Rees}, {Regimbau}, {Rei}, {Reid},
  {Reid}, {Reitze}, {Relton}, {Renzini}, {Rettegno}, {Rezac}, {Ricci},
  {Richards}, {Richardson}, {Richardson}, {Riemenschneider}, {Riles},
  {Rinaldi}, {Rink}, {Rizzo}, {Robertson}, {Robie}, {Robinet}, {Rocchi},
  {Rodriguez}, {Rolland}, {Rollins}, {Romanelli}, {Romano}, {Romel},
  {Romero-Rodr{\'\i}guez}, {Romero-Shaw}, {Romie}, {Ronchini}, {Rosa}, {Rose},
  {Rosi{\'n}ska}, {Ross}, {Rowan}, {Rowlinson}, {Roy}, {Roy}, {Roy}, {Rozza},
  {Ruggi}, {Ryan}, {Sachdev}, {Sadecki}, {Sadiq}, {Sago}, {Saito}, {Saito},
  {Sakai}, {Sakai}, {Sakellariadou}, {Sakuno}, {Salafia}, {Salconi}, {Saleem},
  {Salemi}, {Samajdar}, {Sanchez}, {Sanchez}, {Sanchez}, {Sanchis-Gual},
  {Sanders}, {Sanuy}, {Saravanan}, {Sarin}, {Sassolas}, {Satari},
  {Sathyaprakash}, {Sato}, {Sato}, {Sauter}, {Savage}, {Sawada}, {Sawant},
  {Sawant}, {Sayah}, {Schaetzl}, {Scheel}, {Scheuer}, {Schiworski}, {Schmidt},
  {Schmidt}, {Schnabel}, {Schneewind}, {Schofield}, {Sch{\"o}nbeck}, {Schulte},
  {Schutz}, {Schwartz}, {Scott}, {Scott}, {Seglar-Arroyo}, {Sekiguchi},
  {Sekiguchi}, {Sellers}, {Sengupta}, {Sentenac}, {Seo}, {Sequino}, {Sergeev},
  {Setyawati}, {Shaffer}, {Shahriar}, {Shams}, {Shao}, {Sharma}, {Sharma},
  {Shawhan}, {Shcheblanov}, {Shibagaki}, {Shikauchi}, {Shimizu}, {Shimoda},
  {Shimode}, {Shinkai}, {Shishido}, {Shoda}, {Shoemaker}, {Shoemaker},
  {ShyamSundar}, {Sieniawska}, {Sigg}, {Singer}, {Singh}, {Singh}, {Singha},
  {Sintes}, {Sipala}, {Skliris}, {Slagmolen}, {Slaven-Blair}, {Smetana},
  {Smith}, {Smith}, {Soldateschi}, {Somala}, {Somiya}, {Son}, {Soni}, {Soni},
  {Sordini}, {Sorrentino}, {Sorrentino}, {Sotani}, {Soulard}, {Souradeep},
  {Sowell}, {Spagnuolo}, {Spencer}, {Spera}, {Srinivasan}, {Srivastava},
  {Srivastava}, {Staats}, {Stachie}, {Steer}, {Steinlechner}, {Steinlechner},
  {Stops}, {Stover}, {Strain}, {Strang}, {Stratta}, {Strunk}, {Sturani},
  {Stuver}, {Sudhagar}, {Sudhir}, {Sugimoto}, {Suh}, {Summerscales}, {Sun},
  {Sun}, {Sunil}, {Sur}, {Suresh}, {Sutton}, {Suzuki}, {Suzuki}, {Swinkels},
  {Szczepa{\'n}czyk}, {Szewczyk}, {Tacca}, {Tagoshi}, {Tait}, {Takahashi},
  {Takahashi}, {Takamori}, {Takano}, {Takeda}, {Takeda}, {Talbot}, {Talbot},
  {Tanaka}, {Tanaka}, {Tanaka}, {Tanaka}, {Tanaka}, {Tanasijczuk}, {Tanioka},
  {Tanner}, {Tao}, {Tao}, {Tapia San Mart{\'\i}n}, {Taranto}, {Tasson},
  {Telada}, {Tenorio}, {Terhune}, {Terkowski}, {Thirugnanasambandam}, {Thomas},
  {Thomas}, {Thompson}, {Thondapu}, {Thorne}, {Thrane}, {Tiwari}, {Tiwari},
  {Tiwari}, {Toivonen}, {Toland}, {Tolley}, {Tomaru}, {Tomigami}, {Tomura},
  {Tonelli}, {Torres-Forn{\'e}}, {Torrie}, {Tosta e Melo}, {T{\"o}yr{\"a}},
  {Trapananti}, {Travasso}, {Traylor}, {Trevor}, {Tringali}, {Tripathee},
  {Troiano}, {Trovato}, {Trozzo}, {Trudeau}, {Tsai}, {Tsai}, {Tsang}, {Tsang},
  {Tsao}, {Tse}, {Tso}, {Tsubono}, {Tsuchida}, {Tsukada}, {Tsuna}, {Tsutsui},
  {Tsuzuki}, {Turbang}, {Turconi}, {Tuyenbayev}, {Ubhi}, {Uchikata},
  {Uchiyama}, {Udall}, {Ueda}, {Uehara}, {Ueno}, {Ueshima}, {Unnikrishnan},
  {Uraguchi}, {Urban}, {Ushiba}, {Utina}, {Vahlbruch}, {Vajente}, {Vajpeyi},
  {Valdes}, {Valentini}, {Valsan}, {van Bakel}, {van Beuzekom}, {van den
  Brand}, {Van Den Broeck}, {Vander-Hyde}, {van der Schaaf}, {van Heijningen},
  {Vanosky}, {van Putten}, {van Remortel}, {Vardaro}, {Vargas}, {Varma},
  {Vas{\'u}th}, {Vecchio}, {Vedovato}, {Veitch}, {Veitch}, {Venneberg},
  {Venugopalan}, {Verkindt}, {Verma}, {Verma}, {Veske}, {Vetrano},
  {Vicer{\'e}}, {Vidyant}, {Viets}, {Vijaykumar}, {Villa-Ortega}, {Vinet},
  {Virtuoso}, {Vitale}, {Vo}, {Vocca}, {von Reis}, {von Wrangel}, {Vorvick},
  {Vyatchanin}, {Wade}, {Wade}, {Wagner}, {Walet}, {Walker}, {Wallace},
  {Wallace}, {Walsh}, {Wang}, {Wang}, {Wang}, {Ward}, {Warner}, {Was},
  {Washimi}, {Washington}, {Watchi}, {Weaver}, {Webster}, {Weinert},
  {Weinstein}, {Weiss}, {Weller}, {Wellmann}, {Wen}, {We{\ss}els}, {Wette},
  {Whelan}, {White}, {Whiting}, {Whittle}, {Wilken}, {Williams}, {Williams},
  {Williamson}, {Willis}, {Willke}, {Wilson}, {Winkler}, {Wipf}, {Wlodarczyk},
  {Woan}, {Woehler}, {Wofford}, {Wong}, {Wu}, {Wu}, {Wu}, {Wu}, {Wysocki},
  {Xiao}, {Xu}, {Yamada}, {Yamamoto}, {Yamamoto}, {Yamamoto}, {Yamamoto},
  {Yamashita}, {Yamazaki}, {Yang}, {Yang}, {Yang}, {Yang}, {Yang}, {Yap},
  {Yeeles}, {Yelikar}, {Ying}, {Yokogawa}, {Yokoyama}, {Yokozawa}, {Yoo},
  {Yoshioka}, {Yu}, {Yu}, {Yuzurihara}, {Zadro{\.z}ny}, {Zanolin}, {Zeidler},
  {Zelenova}, {Zendri}, {Zevin}, {Zhan}, {Zhang}, {Zhang}, {Zhang}, {Zhang},
  {Zhang}, {Zhao}, {Zhao}, {Zhao}, {Zhao}, {Zhou}, {Zhou}, {Zhu}, {Zhu},
  {Zimmerman}, {Zlochower}, {Zucker}, \& {Zweizig}}]{LigoO3b}
{The LIGO Scientific Collaboration}, {the Virgo Collaboration}, {the KAGRA
  Collaboration}, {et~al.} 2023, Physical Review X, 13, 011048

\bibitem[{{Thorne}(1974)}]{Thorne0774}
{Thorne}, K.~S. 1974, \apj, 191, 507

\bibitem[{{Tomsick} {et~al.}(2018){Tomsick}, {Parker}, {Garc{\'\i}a},
  {Yamaoka}, {Barret}, {Chiu}, {Clavel}, {Fabian}, {F{\"u}rst}, {Gandhi},
  {Grinberg}, {Miller}, {Pottschmidt}, \& {Walton}}]{Tomsick2018}
{Tomsick}, J.~A., {Parker}, M.~L., {Garc{\'\i}a}, J.~A., {et~al.} 2018, \apj,
  855, 3

\bibitem[{{Tripathi} {et~al.}(2020){Tripathi}, {Zhou}, {Abdikamalov},
  {Ayzenberg}, {Bambi}, {Gou}, {Grinberg}, {Liu}, \& {Steiner}}]{Tripathi0720}
{Tripathi}, A., {Zhou}, M., {Abdikamalov}, A.~B., {et~al.} 2020, \apj, 897, 84

\bibitem[{{Walton} {et~al.}(2016){Walton}, {Tomsick}, {Madsen}, {Grinberg},
  {Barret}, {Boggs}, {Christensen}, {Clavel}, {Craig}, {Fabian}, {Fuerst},
  {Hailey}, {Harrison}, {Miller}, {Parker}, {Rahoui}, {Stern}, {Tao}, {Wilms},
  \& {Zhang}}]{wal16}
{Walton}, D.~J., {Tomsick}, J.~A., {Madsen}, K.~K., {et~al.} 2016, \apj, 826,
  87

\bibitem[{{Wiktorowicz} {et~al.}(2014){Wiktorowicz}, {Belczynski}, \&
  {Maccarone}}]{Wiktorowicz0914}
{Wiktorowicz}, G., {Belczynski}, K., \& {Maccarone}, T. 2014, in Binary
  Systems, their Evolution and Environments, ed. R.~{de Grijs}, 37

\bibitem[{{Wiktorowicz} {et~al.}(2021){Wiktorowicz}, {Lasota}, {Belczynski},
  {Lu}, {Liu}, \& {I{\l}kiewicz}}]{Wiktorowicz0721}
{Wiktorowicz}, G., {Lasota}, J.-P., {Belczynski}, K., {et~al.} 2021, \apj, 918,
  60

\bibitem[{{Yoshioka} {et~al.}(2022){Yoshioka}, {Mineshige}, {Ohsuga},
  {Kawashima}, \& {Kitaki}}]{Yoshioka1222}
{Yoshioka}, S., {Mineshige}, S., {Ohsuga}, K., {Kawashima}, T., \& {Kitaki}, T.
  2022, \pasj, 74, 1378

\bibitem[{{Zdziarski} {et~al.}(2024{\natexlab{a}}){Zdziarski}, {Banerjee},
  {Chand}, {Dewangan}, {Misra}, {Szanecki}, \&
  {Nied{\'z}wiecki}}]{Zdziarski0224a}
{Zdziarski}, A.~A., {Banerjee}, S., {Chand}, S., {et~al.} 2024{\natexlab{a}},
  \apj, 962, 101

\bibitem[{{Zdziarski} {et~al.}(2024{\natexlab{b}}){Zdziarski}, {Chand},
  {Banerjee}, {Szanecki}, {Janiuk}, {Lubi{\'n}ski}, {Nied{\'z}wiecki},
  {Dewangan}, \& {Misra}}]{Zdziarski0224b}
{Zdziarski}, A.~A., {Chand}, S., {Banerjee}, S., {et~al.} 2024{\natexlab{b}},
  \apjl, 967, L9

\bibitem[{{Zdziarski} {et~al.}(1996){Zdziarski}, {Johnson}, \&
  {Magdziarz}}]{Zdziarski99}
{Zdziarski}, A.~A., {Johnson}, W.~N., \& {Magdziarz}, P. 1996, \mnras, 283, 193

\bibitem[{{Zdziarski} {et~al.}(2020){Zdziarski}, {Szanecki}, {Poutanen},
  {Gierli{\'n}ski}, \& {Biernacki}}]{Zdziarski20}
{Zdziarski}, A.~A., {Szanecki}, M., {Poutanen}, J., {Gierli{\'n}ski}, M., \&
  {Biernacki}, P. 2020, \mnras, 492, 5234

\bibitem[{{Zdziarski} {et~al.}(2023){Zdziarski}, {Veledina}, {Szanecki},
  {Green}, {Bright}, \& {Williams}}]{Zdziarski2023a}
{Zdziarski}, A.~A., {Veledina}, A., {Szanecki}, M., {et~al.} 2023, \apjl, 951,
  L45

\bibitem[{{Zhao} {et~al.}(2021{\natexlab{a}}){Zhao}, {Gou}, {Dong}, {Tuo},
  {Liao}, {Li}, {Jia}, {Feng}, \& {Steiner}}]{Zhao0821}
{Zhao}, X., {Gou}, L., {Dong}, Y., {et~al.} 2021{\natexlab{a}}, \apj, 916, 108

\bibitem[{{Zhao} {et~al.}(2021{\natexlab{b}}){Zhao}, {Gou}, {Dong}, {Zheng},
  {Steiner}, {Miller-Jones}, {Bahramian}, {Orosz}, \& {Feng}}]{Zhao2021}
{Zhao}, X., {Gou}, L., {Dong}, Y., {et~al.} 2021{\natexlab{b}}, \apj, 908, 117

\end{thebibliography}
\begin{appendix}

\section{Fitting Cyg X-1 spectra}
\label{app.1}

We fit Cyg X-1 using two models: 1) considering a standard disc and 2) where the disc is 
covered by some warm optically thick material (models a and b, respectively). In both cases we model 
the spectra using {\sc xspec} v.12.13.0 \citep{Arnaud96}.

For the standard disc we use the {\sc xspec} model {\sc kerrbb} \citep{Li05}. This considers 
multi-colour blackbody emission originating from a disc structure, assuming \citet{nt73} emissivity. 
All relativistic effects are taken into account, including ray-tracing from the disc to the observer. 
Additionally, this model always assumes that the disc extends to the ISCO.

However, Cyg X-1 always shows a high energy tail that cannot be disc emission. This is generally 
understood as originating from a hot optically thin inner corona, which Compton up-scatters incident 
disc photons to higher energies. To model this we convolve our disc emission, {\sc kerrbb}, with 
{\sc simpl} \citep{Steiner09}. This takes an input seed spectrum (in this case {\sc kerrbb}) and 
scatters a fraction of the seed photons into a power-law component (hence emulating a Comptonised 
spectrum). In {\sc xspec} syntax, the model is {\sc simpl*kerrbb}.

For model b we also have the addition of a warm optically thick skin covering the disc. This will 
also Compton up-scatter the disc photon to higher energies, but not sufficiently so to give a high 
energy tail. Instead this will give a spectrum that looks similar to a disc, but with the peak 
shifted to slightly higher energies and a shallower drop-off than the usual Wien tail. We model 
this using the convolution model {\sc thcomp} \citep{Zdziarski99, Zdziarski20}. This takes an input 
seed spectrum and Compton scatters to higher energies, similar to {\sc simpl}, however unlike 
{\sc simpl} {\sc thcomp} takes the electron temperature of the Comptonising plasma as an input 
argument, setting the high energy turn over. Additionally, for this optically thick skin we assume 
that all input seed photons are Compton scattered, unlike the case for the high-energy tail where 
only a fraction are scattered. In {\sc xspec} syntax the model is now {\sc csimpl*thcomp*kerrbb}.

The line of sight absorption to Cyg X-1 is complex and variable. Hence we need to include an 
absorption component in both our models, for which we use {\sc tbabs}. Due to the variable nature 
of the absorber in Cyg X-1 we leave the column-density in {\sc tbabs} as a free parameter 
throughout. We note that in Fig. \ref{fig:cyg_3panel} 
we show the spectrum corrected for absorption in colour and the original uncorrected spectrum in 
grey.

As the data originate from two different instruments, XIS (soft-xrays) and HXD (hard-xrays), 
while both onboard Suzaku, we included a cross-calibration constant in all our models. Throughout, 
this was fixed to 1.15, that is:\ increasing the model flux in the hard band by a factor 1.15.
Finally, the total {\sc xspec} models are: {\sc const*tbabs*csimpl*kerrbb} for model a, and 
{\sc const*tbabs*csimpl*thcomp*kerrbb} for model b.

\section{Revisiting M33-X7}
\label{app.2}

\begin{figure*}
    \centering
    \includegraphics[height=10.0 cm, width=\textwidth]{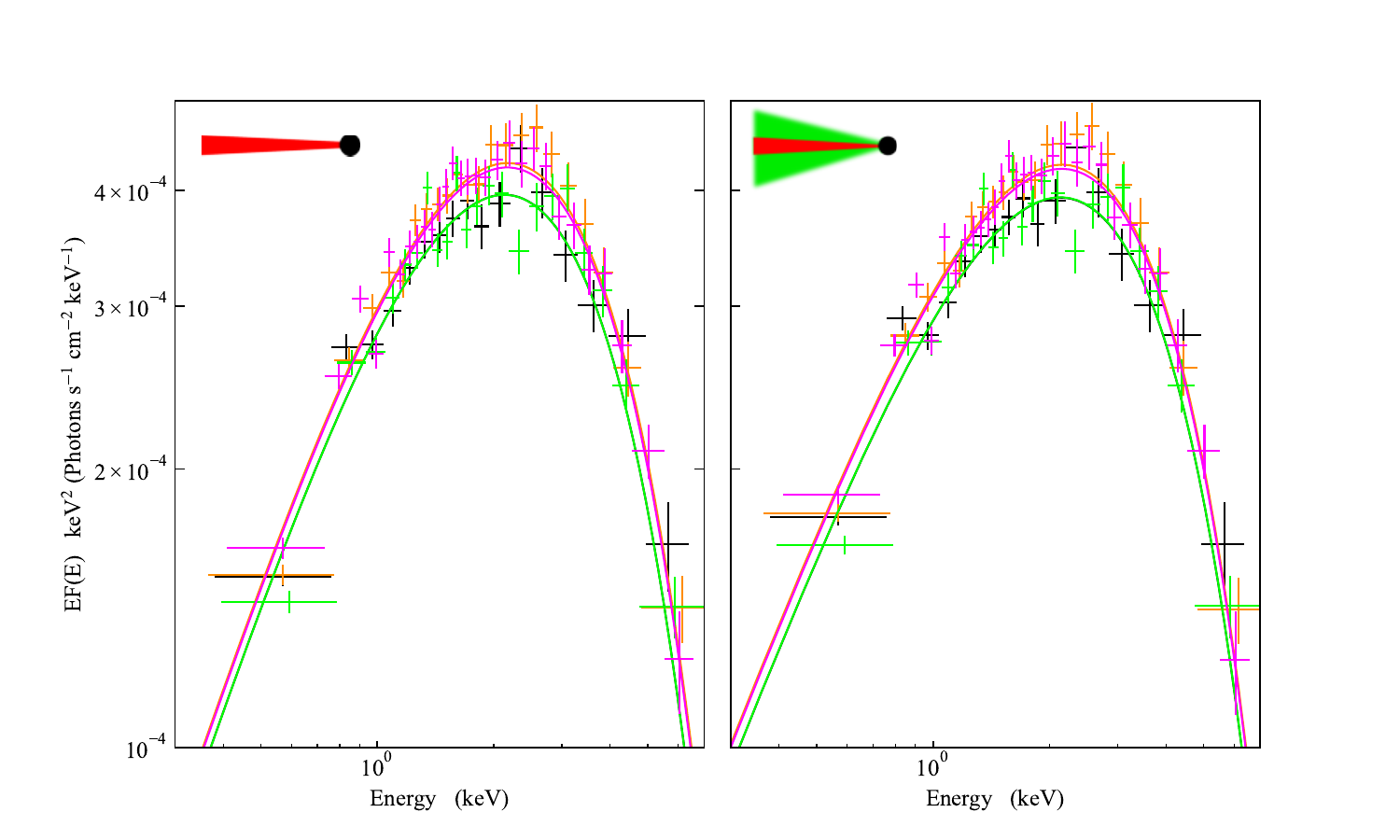}
    \caption{\textit{Chandra} ACIS spectra of M33-X7, re-binned for clarity. We have used the four 'gold' spectra from \citet{Liu2008}, corresponding to obs IDs: acisf1730 (black), acisf6382 (orange), acisf6387 (green), and acisf6376 (magenta). The solid lines show the best fitting model to each spectrum, where we have used a standard disc ({\sc kerrbb}) on the left, and a disc covered by a warm Comptonising medium ({\sc thcomp*kerrbb}) on the right. The magenta line corresponds to the mass $15.65\,\msun$ (\citealt{Orosz2007}, the green line to the mass fixed to $11.4\,\msun$, derived in \citet{Ramachandran2022}. We note that every spectrum has been de-absorbed using the best fit model. For each model the spectra were fit simultaneously, with only the mass accretion rate, $\dot{M}$, allowed to vary between them.}
    \label{fig:m33_spec}
\end{figure*}

\begin{figure*}
    \centering
    \includegraphics[height=10.0 cm, width=\textwidth]{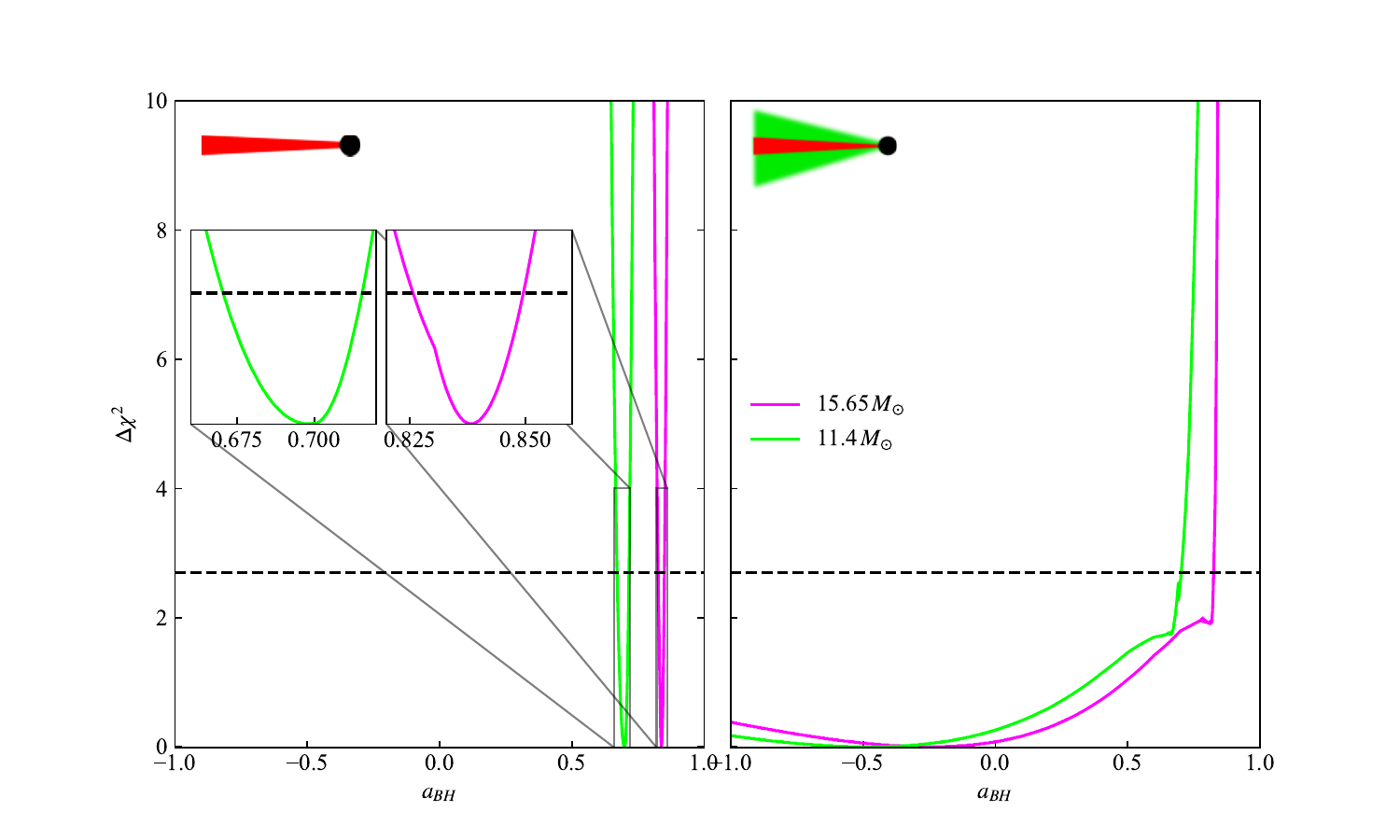}
    \caption{$\Delta \chi^{2}$ curves with black hole spin as the interesting parameter for our spectral fits to M33-X7. The left panel shows the case for a standard accretion disc model ({\sc kerrbb}), while the right panel shows the case where there is a warm Comptonising material above the disc ({\sc thcomp*kerrbb}). We have included both the previous mass estimate of $15.65\,\msun$ (\citealt{Orosz2007}, magenta line) and the updated estimate of $11.4\,\msun$ (\citealt{Ramachandran2022}, green line). The dashed black horizontal line shows $\Delta \chi^{2}=2.7$, indicating $90\,\%$ confidence for one interesting parameter.}
    \label{fig:m33_chisq}
\end{figure*}

Here we revisit the black hole spin estimate of M33-X7, performed by \citet{Liu2008}, using the same methodology as for Cyg X-1. We extract the four \textit{Chandra} ACIS spectra, referred to as the 'gold' spectra in \citet{Liu2008}. An initial inspection of the data show strongly disc dominated spectra, with no clear non-thermal high energy tail within the observed energy range. Hence, we used a slightly simpler model than with Cyg X-1, modelling the continuum as either a pure disc (using {\sc kerrbb}, \citet{Li05}) or a disc with a warm Comptonising medium above it (using {\sc thcomp*kerrbb}, \citealt{Zdziarski99, Zdziarski20}). We also include the effects of absorption along our line of sight, which will affect the low energy emission, using {\sc tbabs}. Hence the total {\sc xspec} models are: {\sc tbabs*kerrbb} for the pure disc case, and {\sc tbabs*thcomp*kerrbb} for the warm Comptonised case.

We note that since the analysis of \citet{Liu2008} the black hole mass of M33 X-7 has been updated from $15.65\,\msun$ \citep{Orosz2007} to $11.4\,\msun$ \citep{Ramachandran2022}. The main effect of this will be to lower the black hole spin estimate in the case of a pure disc, as a lower black hole mass will shift the peak of the spectrum to higher energies, much in the same way as an increase in black hole spin will do. For completeness we fit the data using both the new and previous mass estimates, to highlight that our conclusions are not sensitive to this change.

Figure \ref{fig:m33_spec} shows the four ACIS spectra along with their respective model fits, for the pure disc (left) and warm Comptonised medium (right). The data show a change in normalisation between some of the spectra, corresponding to a $\sim 10\,\%$ change in the mass-accretion rate (see e.g the difference between the green-black data model and the magenta-orange data model in Fig. \ref{fig:m33_spec}). Hence, during the spectral fitting we allow the mass-accretion rate to vary between the datasets, but then tie all other parameters as these should not vary between observations. The spectra were grouped such that each energy bin contained a minimum of 20 counts, allowing for the use of $\chi^{2}$ statistics.

Figure \ref{fig:m33_chisq} shows the change in $\chi^{2}$ as we step the fit through the possible values for black hole spin, with a standard disc on the left and a disc covered by a warm Comptonising material on the right. The standard disc clearly gives a strongly constrained value for the black hole spin, and in fact for the old mass estimate we recover the value derived in \citet{Liu2008} of $\sim 0.84$. However, it is clear that a change in mass will affect this result, with the new mass estimate reducing the black hole spin to $\sim 0.70$. However, the accretion flow is not necessarily characterised by a standard disc. Instead it is likely that it could be covered in a warm Comptonising material, as discussed previously for Cyg X-1 and in LMC X-1 \citep{Zdziarski0224a, Zdziarski0224b}. Unlike the case of a pure disc, this gives completely unconstrained black hole spin, with acceptable values ranging from -1 to $\sim 0.9$ at $90\,\%$ confidence. As with both Cyg X-1 and LMC X-1, this demonstrates the strongly model dependent nature of black hole spin estimates.

\end{appendix}

\end{document}